\documentstyle[preprint,aps]{revtex}
\tighten 
\begin{document}
\title{Neutrinos in Strong Magnetic Fields}
\author{A. P\'erez Mart\'{\i}nez$^{1, 2}$, 
A. Am\'ezaga Hechavarr\'{\i}a$^{1}$, D. Oliva Ag\"uero$^{1}$, 
 { H. P\'erez Rojas}$^{1,3}$ and
{S. Rodr\'{\i}guez Romo}$^{2}$  \\ \vskip 1.5cm
$^1$ ICIMAF, Calle E No. 309,  10400 La Habana, Cuba \\
$^2$ Centro de Investigaciones Te\'oricas, FES-Cuautitl\'an, UNAM, 
 M\'exico. \\
$^3$ CINVESTAV-IPN, A.P. 14-740, 07000 M\'exico, D.F.\\
 }
\maketitle

\begin{abstract}
We compute the dispersion curves for neutrinos propagating  in an
extremely dense electroweak plasma,
in the presence of very strong magnetic fields of order $B \le M_W^2/e$. 
The neutrino self-energy is calculated in the one-loop approximation.
We consider only contributions of the first Landau level to the
propagator of the $W$-bosons, and distinguish between motion parallel or
perpendicular to the external magnetic field.
We find that the neutrino dispersion curve for parallel propagation to the
field suggests a superfluid behavior. An interesting analogy with
fractional QHE is pointed out. We obtain a neutrino effective mass which
increases with the magnetic field.
\end{abstract}

\section{Introduction}
The present paper is   addressed to investigate the dispersion equation of
massless
neutral fermions, interacting with charged fermions and massive vector
bosons, propagating in a medium at finite temperature and density, and in
presence of an extremely strong magnetic field.

The propagation of neutrinos in an electroweak plasma has been 
studied and the dispersion equation for the quasiparticles was obtained
\cite{yo,a,elm1}. The  spectrum found exhibits, in some extreme 
conditions, a superfluid behavior.  

In the present paper we consider the role of extremely strong magnetic
fields
as a possible mechanism for generating an effective neutrino mass in a
very dense medium. We find 
again a superfluid behavior for the neutrinos moving parallel to the
external magnetic field, provided it is 
strong enough.

The propagation of neutrinos in magnetized
media, assuming no  dependence of the $W$-propagator   
on the magnetic field, has been done by computing  two
types of
diagrams (bubble and tadpole) \cite{elm}. We work with the effective
action,
the generating functional of irreducible one-particle inverse Green's
functions.
Thus,  in the evaluation of  the inverse Green's function, whose
zeros give the dispersion equation, the tadpole diagrams do not appear.   
The tadpole
diagram is reducible and does not contribute to it.

For strong enough magnetic fields,  a gas of charged bosons  undergoes 
Bose-Einstein condensation \cite{conden}. This suggests that  it is the 
ground state of the bosons which  play the main role.
In our calculation we will take into account only the part of the $W$ 
propagator which contains the ground state energy \cite{polon}. When the
momentum is small and the magnetic field is high enough $eB\le M_{W}^2$, 
so that the term $1/\sqrt{M_{W}^2-eB}$ dominates, the main contribution 
to the propagator comes from the low momentum gauge bosons ($W$-condensate)
\cite{polon,conden}.

The expressions we derive for the neutrino dispersion equation  are valid 
for strong fields, close to the limiting value $M^2_{W}/e$. Such fields
are at present conjectured to exist in the cores of neutron stars
\cite{Chakrabarty}, and they may 
have also existed in the early universe, in which case  the observed
galactic 
and intergalactic
magnetic fields are viewed as relics of huge primordial fields \cite{29}-
\cite{33}.
  However, on a more general basis, our
results apply to any  massless fermions interacting with
vector bosons of mass $M$ in strong magnetic fields $eB\le M^2$.

  The first part of this paper (section 2)  presents some 
general expressions of Green functions  in a magnetized medium, which will be used  subsequently. In section 3, 
we find the  mass operator in the limit of high magnetic field, 
after performing the sum over $p_{4}$. These results are very general and
apply equally well to   the limiting cases of high temperature and
high density. In section 4, we analyze the limit $T \to 0$, 
corresponding to a degenerate fermion gas (large fermion density). We also
consider the role of $W$-boson condensation. In section 5, the dispersion
equations are discussed. We analyze two cases of neutrino propagation,
parallel and perpendicular to the magnetic field.
For parallel motion,  an effective mass results only in the sense of $\lim_{k_{3}\to 0}\omega\neq 0$. 
When the motion is perpendicular to the magnetic field, a minimum appears in
the dispersion equation at non-zero momentum and it is possible to define an effective mass in
the more traditional sense of being a local minimum of the dispersion
equation.

Our results depend linearly on the
magnetic field and, in the degenerate case, also linearly on the chemical
potential $\mu$. Other  studies of the mass operator in QED in presence of a 
magnetic field \cite{elm1} analyze different limits: high
temperature, high density, and high magnetic field.
The first two   show  similar dispersion equations \cite{pi} if one substitutes $M=e^2T^2/8$ with $M=e^2\mu^2/(8\pi^2)$.

\section{Neutrino Self-Energy: General Expressions}
\noindent

The neutrino two-point inverse Green function in presence of a magnetic
field reads as
\begin{equation}
S_{\nu}^{-1}(k^*)=P_R(-i\gamma_{\mu}k_{\mu}^* +\Sigma^{W}(k^*))P_L.
\label{ee}
\end{equation}
\noindent
where $k_{\lambda}^{*}=k_{\lambda}-i\mu_{\nu}\delta_{4,\lambda}$, and $P_R=(1+\gamma_{5})/2$ and $P_L=(1-\gamma_{5})/2$ are the
left- and right-handed projection operators. We recall \cite{yo} that the
following relation holds among the chemical potentials $\mu_{\nu} =
\mu_e - \mu_{W^-} $. 
Note  that, in equation (\ref{ee}),  the self-energy
  of the $Z$-boson is not included. It can be neglected since it does
not interact with the magnetic field (at the one-loop level); it is of
order $g'^2/M_Z^2$, whereas the $W$ term is of order $g^2 eB/M_W^2$.

The expression for the self-energy $\Sigma^{W}$ is the following, in configuration space:
\begin{equation}
\Sigma^{W}(x,x^{\prime})=-i\frac{g^2}{2\pi^3}\gamma_{\mu}G^{e}(x,x^{\prime})D^{W}_{\mu\nu}(x-x^{\prime})\gamma_{\nu},  \label{sig}
\end{equation}
It represents the self-energy due to electron-W polarization.

In Euclidean space and in the gauge $A_{\mu}=(0,Bx,0,0)$,  the propagator of the electron is 
\[
G^e(x,x^{\prime})=-\frac{1}{(2\pi^2)}\sum_{n=0}^{\infty}\sum_{p_{4}^{*}}
\int\frac{dp_{3}dp_{2}} {\beta(2\pi)^3(p_{4}^{*^2}+p_{3}^2+m_{e}^2+2eBn)} 
\]
\[
\cdot\left\{ \left( (ip_4-\mu )\gamma _4+ip_3\gamma _3-m_{e})(\sigma _{+}\psi _n\psi
_n+\sigma _{-}\psi _{n-1}\psi _{n-1}\right) \right. 
\]
\begin{equation}
+1/2\sqrt{2eBn}[\gamma _{+}\psi _n\psi _{n-1}-\left. \gamma _{-}\psi
_{n-1}\psi _n]\right\}  \label{ele}
\end{equation}
\[
\cdot{\rm exp}[ip_{4}^{*}(x_{4}-x^{\prime}_{4})+ip_{3}(x_{3}-x^{\prime}_{3})
+ip_{2}(x_{2}-x^{^{\prime}}_{2})],
\]
\noindent
where $\xi=\sqrt{eB}(x_{1}+x_{o})$, $\xi^{\prime}=\sqrt{eB}(x_{1}+x_{o})$,
$x_{o}=\frac{p_{2}}{eB}$, $\sigma ^{\pm }=1/2[1\pm \sigma _z]$,
$\gamma _{\pm}=1/2[\gamma _1\pm i\gamma _2]$,   
$\sigma _3 =i/2[\gamma _1,\gamma _2]$, and  $p_{\lambda}^{*}=p_{\lambda}-i\mu_{W}\delta_{4,\lambda}$.

The W-propagator in a magnetic field has the form 
\begin{equation}
D_{\mu\nu}^W(x,x^{^{\prime}})=\frac{1}{(2(2\pi)^2\beta)} \int dp_{2}dp_{3}
[\frac{R^{-}+R^{+}}{2}\Psi^{1}_{\mu\nu}+R^0\Psi^{2}_{\mu\nu}+
i\frac{(R^{-}-R^{+})}{2}\Psi^{3}_{\mu\nu})]  \label{dmu}
\end{equation}
\[
\cdot \psi_{n}(\xi)\psi_{n}(\xi^{\prime}){\rm exp}[ip_{4}^{*}(x_{4}-x^{\prime}_{4})
+ip_{3}(x_{3}-x^{\prime}_{3}) +ip_{2}(x_{2}-x^{\prime}_{2})]. 
\]
\noindent
where $R^{\pm}=[p_{4}^{*^2} + E_{n^{\prime}}^2 \pm 2eB]^{-1}$, 
 $R^0=[p_{4}^{*2} + E_{n^{\prime}}^2]^{-1}$, 
with  $E_{W}^{2} = M_{W}^2 + p_{3}^2 +2eB(n+1/2)$; 
$\Psi^{1}_{\mu\nu}=\frac{1}{B^2} G^{0 2}_{\mu\nu}$, 
$\Psi^{2}_{\mu\nu}=\delta_{\mu\nu}-\frac{1}{B^2} G^{0 2}_{\mu\nu}$ and 
$\Psi^{3}_{\mu\nu}=\frac{1}{B} G^{0}_{\mu\nu}$ ($G_{\mu\nu}^{02}$ is the
field tensor of the SU(2)xU(1) electromagnetic external field). Concerning
the gauge fixing term, we are taking $D^W_{\mu\nu}$ in a transverse gauge
which is expected to guarantee the gauge independence of the neutrino
spectrum.

The poles of $D_{\mu\nu}^W$ are located at 
\begin{equation}
E^W_g =E^W_{-1} = \sqrt{p_3^2 + m_W^2 - e B},  \label{gs}
\end{equation}
which is the ground state energy, and at
\begin{equation}
E^W_n =  \sqrt{p_3^2 + m_W^2 +2eB(n + \frac{1}{2})},
\end{equation}
where $n = 0,1,2,\ldots$  with  degeneracy $\beta_n = 3 - \delta_{0n}$.
The ground state energy (\ref{gs}) is unstable for $p_3^2 < eB - M_W^2$.
The
analog of the Euler-Heisenberg vacuum energy due to vector boson
polarization is \cite{polon}
\begin{equation}
U_{W} = -\frac{1}{16\pi^2}\int_{0}^{\infty}\frac{dt}{t^2}%
e^{-M_{W}^2t}[eB\,{\rm csch}(eBt)(1+2\,{\rm cosh} 2eBt)
-\frac{3}{t}-\frac{7}{2}e^2B^2t].
\end{equation}
Convergence of this expression is only possible for $eB < M_W^2$,  i.e.
the vacuum becomes unstable for
$eB \geq M_W^2 $.
This problem has been the subject of investigation mainly by Nielsen,
Olesen and
Ambjorn \cite{Olesen},\cite{Ambjorn}. In the last reference, a static
magnetic solution of classical electroweak equations, corresponding to a
vacuum condensate of $W$ and $Z$ bosons, is found. It is valid above
the critical value $B_c = M_W^2/e$. The vacuum bears  the properties
of a ferromagnet or an anti-screening superconductor.

We are interested in the Fourier transform of (\ref{sig}). It
requires  rather long calculations involving the Fourier
transform for two  Hermite functions, which lead to functions of
generalized Laguerre polynomials. Eventually we find 
\begin{equation}
\Sigma^{W}(k)= \frac{g^2}{2\pi^2}(\sum_{p_{4}^{*}}\int\frac{dp_{3}}{(2\pi)^2}%
G^{e}(p_{3}+k_{3},p_{4}^{*}+k_{4}^{*},n^{\prime}) \Sigma_{\alpha\beta})P_{L},  
\label{sig1}
\end{equation}
where
\[G _{e}(p_3+k_3,p_4+k_4,n^{\prime})= \left( {(p_3+k_3)^2+(p_4+k_4)^2+m_{e}^2+2eBn'}  \right)^{-1} \]
and $\Sigma_{\alpha\beta}$ is a $4\times4$ matrix whose elements are the
following ($\Sigma_{12}=\Sigma_{21}=\Sigma_{34}=\Sigma_{43}=0$):
\[ \Sigma_{11}=B_{1}(i(p_{4}^{*}+k_{4}^{*}))T^{*}_{n^{\prime}-1,n}T_{n^{%
\prime}-1,n}+B_1(-i(p_{4}^{*}+k_{4}^{*}))
T^{*}_{n^{\prime}-1,n}T_{n^{\prime}-1,n}  \]
\[ +2A_{1}(i(p_{4}^{*}+k_{4}^{*}))T^{*}_{n^{\prime},n}T_{n^{\prime},n},  \]
\[ \Sigma_{13}=-2A_1(p_{3}+k_{3})T^{*}_{n^{\prime},n}T_{n^{\prime},n},  \]
\[ \Sigma_{14}=-2iB_1\sqrt{2eBn}T^{*}_{n^{\prime}-1,n}T_{n^{\prime},n}- 2iC%
\sqrt{2eBn}T^{*}_{n^{\prime},n}T_{n^{\prime}-1,n},  \]
\[ \Sigma_{22}=B_1(i(p_{4}^{*}+k_{4}^{*}))T^{*}_{n^{\prime},n}T^{*}_{n^{%
\prime},n}+B_1(-i(p_{4}^{*}+k_{4}^{*}))T^{*}_{n',n}T_{n',n}  \]
\[ +2A_{1}(i(p_{4}^{*}+k_{4}^{*}))T^{*}_{n'-1,n}T_{n'-1,n},  \]
\[ \Sigma_{23}=-2iB_1\sqrt{2eBn}T^{*}_{n^{\prime},n}T_{n^{\prime},n}-2iB_1\sqrt{%
2eBn}T^{*}_{n^{\prime}-1,n}T_{n^{\prime},n} -2iC\sqrt{2eBn}%
T^{*}_{n^{\prime}-1,n}T_{n^{\prime},n},  \]
\[ \Sigma_{24}=2A_{1}(p_{3}+k_{3})T^{*}_{n^{\prime}-1,n}T_{n^{\prime}-1,n},  \]
\[ \Sigma_{31}=2A_{1}(p_{3}+k_{3})T^{*}_{n^{\prime},n}T_{n^{\prime},n},  \]
\[ \Sigma_{32}=2iB_1\sqrt{2eBn}T^{*}_{n^{\prime}-1,n}T_{n^{\prime},n}+2iC\sqrt{%
2eBn}T^{*}_{n^{\prime}-1,n}T_{n^{\prime},n},  \]
\[ \Sigma_{33}=B_1(i(p_{4}^{*}+k_{4}^{*}))T^{*}_{n^{\prime}-1,n}T_{n^{%
\prime}-1,n}+B_1(-i(p_{4}^{*}+k_{4}^{*}))T^{*}_{n^{\prime}-1,n}T_{n^{%
\prime}-1,n}  \]
\[ 2A_{1}(-i(p_{4}^{*}+k_{4}^{*}))T^{*}_{n^{\prime},n}T_{n^{\prime},n},  \]
\[ \Sigma_{42}=2A_{1}(p_{3}+k_{3})T^{*}_{n^{\prime}-1,n}T_{n^{\prime}-1,n},  \]
\[ \Sigma_{44}=B_1(i(p_{4}^{*}+k_{4}^{*}))T^{*}_{n^{\prime},n}T_{n^{%
\prime},n}+B_1(-i(p_{4}^{*}+k_{4}^{*}))T^{*}_{n^{\prime},n}T_{n^{\prime},n}  \]
\[ -2A_{1}(-i(p_{4}^{*}+k_{4}^{*}))T^{*}_{n^{\prime}-1,n}T_{n^{\prime}-1,n},  \]
\noindent
where
$A_1= {-(R^{+}+R^{-})}/{2}+2R^0$, $ B_1=R^0$, $C=- {i}/{2}(R^{+}-R^{-})$, 
\[T_{n,m}=\left(\frac{n!}{m!}\right)^{1/2}\left(\frac{k_1+ik_2}{2}\right)^{m-n}e^{-i\frac{k_1k_2}{2}-\frac{k_1^2+k_2^2}{4}}
L_{m}^{m-n}((k_1^2+k_2^2)/2),\]
and $L_{m}^{m-n}$ are the Laguerre polynomials.

\section{Neutrino Self-Energy in the High Magnetic Field Limit}
Let us now consider the limit of an extremely strong magnetic field ($eB$ close to, but smaller than, $M_W^2$).  The $W$ propagator is dominated by the Landau ground
state term ($n=0$), which   means we keep in (\ref{dmu}) 
only  terms proportional to $R^{-}$.

Furthermore, the condition $eB>\mu_e^2-m^2$ implies that the
only electron state which contributes to the mass operator is $n^{\prime}=0$. 
The sum $\sum_{n^{\prime}=0}^{\infty}$ can be approximated to 
$\sum_{n^{\prime}=0}^{n_{\mu}}$, where $n_{\mu}$ is the integer part 
of $ ({\mu_e^2-m^2})/{2eB}$,
which is zero whenever $\mu_e^2 < eB$, i.e., for most   cases
of interest. Hence, we   concentrate on the case when both electron and
$W$-boson are in the Landau ground state, and their quantum numbers are 
$n=n^{\prime}=0$.

Keeping in mind the above approximations, equation  (\ref{sig1}) becomes
\begin{equation}
\Sigma^{W}(k)=\frac{g^2eB}{(2\pi)^2}(\sum_{p_{4}}\int dp_{3}
G^{o}_{e}(p_3+k_3,p_4+k_4)\Sigma_{\alpha\beta})P_{L},  \label{mag}
\end{equation}
with $\Sigma_{\alpha\beta}=R^{-}(p_3,p_4)e^{-k_{\perp}^2/eB}\Sigma^{\prime}_{\alpha\beta}$, 
and $R^{-}_{o}=[M_{W}^2 +p_{4}^{*2} +p_{3}^2 -2eB]^{-1}$, 
$G^{e}(p_{3}+k_{3}) ^{-1}=\left[ {(p_{4}^{*}+k_{4}^{*})^2 +(p_{3}+k_{3})^2 +m_{e}^2}\right]^{-1}$. 
The matrix $\Sigma^{\prime}_{\alpha\beta}$ simplifies and takes  the form
\begin{equation}
\Sigma^{\prime}_{\alpha\beta}=\left ( 
\begin{array}{lccr}
(i(p^{*}_{4}+k^{*}_{4}) & 0 & (p_{3}+k_{3}) & 0 \\ 
0 & 0 & 0 & 0 \\ 
-(p_{3}+k_{3}) & 0 & -i(p^{*}_{4}+k^{*}_{4}) & 0 \\ 
0 & 0 & 0 & 0
\end{array}
\right),
\end{equation}

After performing in (\ref{mag}) the sum over $p_{4}^{*}$ and taking the 
analytic continuation ($k_4^{*}\to ik_{o}$) we get a function of the new 
variable $k_{o}-\mu_{\nu}$. The singularities in this variable lead
to the gauge-invariant, physically relevant spectrum. We have 
\[
-i(k_{4}-i\mu_{\nu})=(k_{o}-\mu_{\nu})=\omega+i\Gamma
\]
\noindent
where $\omega$ is the energy and $\Gamma$ the inverse lifetime of the 
neutrino quasiparticles.

We obtain the following  expression for $\Sigma^{W}$: \noindent
\begin{eqnarray}
\Sigma^W_{11}=\frac{g^2eB}{(2\pi)^2}[\omega I_{2}-I_{1}+I_{3}],
\nonumber \\
\Sigma^W_{33}=-\Sigma^{W}_{11}\nonumber\\
\Sigma^W_{13}=-\Sigma^W_{31}=\frac{g^2eB}{(2\pi)^2}[-k_{3}I_{2}+I_{4}-I_{5}], 
\label{element}
\end{eqnarray}
where the integrals $I_{i}$  can be written as (neglecting the vacuum
contributions),
\begin{eqnarray}
I_{1}=\int\frac{dp_{3}}{2Q}\left((J_{oo}-2p_{3}k_{3})(n_{e}-n_{p})
-2\omega E_{e}(n_{e}+n_{p})\right) e^{-k\perp^2/2eB},  \nonumber \\
I_{2}=\int\frac{dp_{3}}{2E_{W}Q}\left((J^{\prime}_{oo}
-2p_{3}k_{3})(n_{W^-}+n_{W^+})+2\omega E_{W}(n_{W}^{-}-n_{W}^{+})\right)
 e^{-k\perp^2/2eB},  \nonumber\\
I_{3}=\int\frac{dp_{3}}{2Q'}\left((J'_{oo}+2p_3k_3)(n_{W^-}-n_{W^+})+
2E_{W}\omega (n_{W}^{-}+n_{W}^{+})\right) e^{-k\perp^2/2eB}, \nonumber\\
I_{4}= \int\frac{p_3dp_{3}}{2E_{e}Q}\left((J_{oo}-2p_{3}k_{3})(n_{e}+n_{p})-2\omega E_{e}
(n_{e}-n_{p})\right) e^{-k\perp^2/2eB},\nonumber \\
I_{5}=\int\frac{p_3dp_{3}}{2E_{W}Q'}\left((J'_{oo}+2p_3k_3)(n_{W^-}+n_{W^+})+
2E_{W}\omega (n_{W}^{-}-n_{W}^{+})\right) e^{-k\perp^2/2eB},
\label{in}
\end{eqnarray}
with
\begin{eqnarray}
Q=(J_{oo}-2p_{3}k_{3})^2-4\omega^2E_{e}^2  \nonumber \\
Q^{\prime}=(J^{\prime}_{oo}+2p_{3}k_{3})^2-4\omega^2E_{W}^2 \nonumber
\end{eqnarray}
and 
\begin{eqnarray}
J_{oo}=z_{1}-eB -m_{e}^2+M_{W}^2,  \nonumber \\
J^{\prime}_{oo}=z_{1}+eB + m_{e}^2-M_{W}^2.  \nonumber
\end{eqnarray}
In the above formulae $E_{e}=\sqrt{((p_{3}+k_{3})^2+m^2_{e})}$ and 
$E_{W}=\sqrt{(p_{3}^2+M_{W}^2-eB)}$ are the Landau ground
state energies of the electron and the $W$-boson, respectively, whereas 
\[
n_{e,p}=[e^{(E_e \mp \mu_e)\beta} + 1]^{-1},\hspace{0.5cm} n_{W^{-},W^{+}}
= [e^{(E_W
\mp \mu_W)\beta} - 1]^{-1}
\]
are respectively the distribution functions of the
electrons, positrons, $W^{-}$ and $W^{+}$  in our plasma.

The mass operator given by the expressions  (\ref{sig1}) is general
and holds also in the high temperature and high density limits. The  branch
points in the denominators $Q$ and $Q^{\prime}$ can be identified 
as thresholds for neutrino absorption in the plasma.  We postpone a careful study of the analytic properties of the mass operator and their implications for the dispersion equation to future work. 
 
\section{Degenerate Case}
In this section we consider the case of degenerate electrons (formally
equivalent to  the limit $T\to0$).
Our results are of interest in the theory of neutron stars as well as in 
the early universe, when  there might
have been magnetic fields $eB\approx M_{W}^2$ ($10^{22}$ gauss).  In the
degenerate case,   the distribution of the electrons is just a step function, $n_{e}=\theta(\mu_{e}-E_{e})$, and there are no positrons left, $n_{p}=0$. Charge neutrality is ensured thanks to some $W^+$ background. From the behavior of the
distributions of $W^{\pm}$    \cite{polon,conden}  it
has been argued that $W$-condensation in the presence of a magnetic field may indeed take place. Thus, the distribution of $W^{+}$ can be approximated by $(2\pi^2)\delta(k_{3})\frac{N_{W}
}{eB}$ ($N_{W}$ is the total density of
$W$-particles in the medium), and for the excited states  $n_{W^+}=0$. 

In this limit, equations (\ref{in}) become 
\begin{eqnarray}
I_1=\frac{g^2eBe^{-k\perp^2/2eB}}{2(2\pi )^2}
\int \frac{dp_3}{2}\frac{\theta (\mu_e-E_e)}
{k_3^2-2p_3k_3-m_e^2-\omega ^2+2\omega E_e+d^2}, \nonumber \\
I_2=\frac{g^2N_We^{-k\perp^2/2eB}}{2(2\pi )^2}\int \frac{dp_3}{(2E_W)}
\frac{\delta (p_3)}{k_3^2+2p_3k_3+m_e^2-\omega ^2-2\omega E_W-d^2},\nonumber \\
I_3=\frac{g^2N_We^{-k\perp^2/2eB}}{2(2\pi )^2}\int \frac{dp_3}{2}
\frac{\delta (p_3)}{k_3^2+2p_3k_3+m_e^2-\omega ^2-2\omega E_W-d^2}, \nonumber\\
I_4=\frac{g^2eBe^{-k\perp^2/2eB}}{2(2\pi) ^2}\int \frac{dp_3}{(2E_e)}
\frac{p_3\theta (\mu_e -E_e)}{k_3^2-2p_3k_3-m_e^2-\omega ^2+2\omega E_e+d^2},  \nonumber \\
I_5=\frac{g^2N_We^{-k\perp^2/2eB}}{2(2\pi) ^2}\int \frac{dp_3}{(2E_W)}\frac{p_3\delta (p_3)}{
k_3^2+2p_3k_3+m_e^2-\omega ^2-2\omega E_W-d^2}, \label{filf} 
\end{eqnarray}
\noindent
where $d=\sqrt{M_{W}^2-eB}$.

\section{Dispersion Equation}
Before solving the dispersion equation, note that we 
work far from the thresholds for neutrino absorption. 
In order to get the dispersion equation we must solve 
\begin{equation}
{\rm det}(-i\gamma_{\mu}k_{\mu}+\Sigma^{W})=0,  \label{dis}
\end{equation}
which yields 
\begin{eqnarray}
&&\left. (k_3^2-\omega ^2)\left[ -\omega ^2-(\Sigma _{11}-\Sigma
_{33})\omega +\Sigma _{11}\Sigma _{33}+(k_3-\Sigma _{13})^2\right] \right.
\label{as} \\
&&\left. +k_{\perp }^2\left[ 2k_3(k_3-\Sigma _{13})+k_{\perp }^2-2\omega
^2-\omega (\Sigma _{11}-\Sigma _{33})\right] =0\right. , 
\label{mim}
\end{eqnarray}

Let us remark that   the limit $eB\to0$ 
 does not mean that $\Sigma^{W}=0$. 
The dispersion equation (\ref{mim}), when $k_{3}\to 0$ $k_{\perp}\to 0$,
leads to a value for $\omega$ different from zero; this value corresponds
to a sort of ``effective mass"\cite{yo} proportional to $eB\mu_{e}/d^2$.
This means that, for fixed electron density, if the magnetic field
grows up to near $M_{W}^2/e$ the``effective mass" grows too. As pointed out
before, the vacuum is unstable for $eB\ge M_{W}^2$.
We shall show below that when the motion is perpendicular to the magnetic
field, the``effective mass'' becomes a  mass in strict sense, since it
is also a minimum of the dispersion curve: in formulas,
$\partial \omega/\partial
k_{\perp}\vert_{k_{\perp}=0}=0$ and $\partial^2 \omega/\partial
k_{\perp}^2\vert_{k_{\perp}=0}>0$.

Since we are considering the degenerate case, the contribution of terms
containing the $W$-boson distribution function can be safely neglected:
for huge magnetic fields,   $N_{W}\approx C << eB$, ($C$ is the 
$W$-condensate), whence only $I_{1}$ and $I_{4}$ contribute in (\ref{mim}).

In order to solve numerically  equation (\ref{mim}), we distinguish 
two cases: motion parallel to the field ($k_{\perp}\to0$) and motion
perpendicular to it ($k_{3}\to 0$). This  equation involves only  $I_{1}$
and  $I_{4}$:
\begin{equation}
(k_{3}^2-\omega^2 + k_{\perp}^2)^2 - (k_{3}^2-\omega^2 +
k_{\perp}^2)(2k_{3}I_{1})
+(k_{3}^2-\omega^2)I^2_{1}=0,  \label{mic}
\end{equation}

\subsection{Motion parallel to magnetic field}
Equation (\ref{mic}) for neutrino propagation along the magnetic field
becomes 
\begin{equation}
(k_{3}^2-\omega^2)^2 +(k_{3}^2-\omega^2)
\left[2I_1^2- 2k_{3}I_1\right]=0.
\label{movpar}
\end{equation}
Figure (1) shows the neutrino dispersion 
curves in this case, having fixed $\mu=100$ $m_{e}$ and $eB=0.9 M^2_{W}$.
It has two branches. One of them corresponds to the light cone. The second one
arises due to the  magnetic field and the finite density. 
The non-zero intercept at  $k_3=0$ is a sort of ``effective mass" 
(for our typical values it is approximately $0.92~m_{e}$).
Besides, the curve has a gap  ($0.68~m_{e}$). 

Interestingly, the curve shows a close analogy to
the collective excitations arising in the fractional quantized Hall
effect. In paper \cite{gir}, a theory of the excitation
spectrum in the fractional Hall effect analogous to Feynman's theory for
the excitation spectrum
of superfluid helium was proposed. A magneto-roton minimum for the
collective excitation spectrum was found, which has a remarkable 
analogy with the minimum obtained in the present case, when the neutrino
propagates parallel to the magnetic field.

It is possible to interpret the gap of the quasiparticle spectrum
as the symptom of a superfluid behavior. A similar interpretation has been 
done in the case of the dispersion of neutrinos in a hot medium without
magnetic field. But here we must observe that the neutrinos interacting with
the $W$-s and electrons must align their spins also along the
magnetic field, leading to weak coupling in pairs, and to condense.

\subsection{Motion perpendicular to the magnetic field}
When  the neutrino moves perpendicularly to the field, we get the following
expression for the dispersion equation
\begin{equation}
-\omega^2(-\omega^2 +2I_{1}^2) 
+k_{\perp}^2(k_{\perp}^2-2\omega^2)=0.\label{min1}
\end{equation}
The numerical solution   also yields two branches (figure 2); one of them gives the same ``effective mass'' as in the parallel case. This   is a mass in
the proper sense since it is the minimum of the dispersion equation.  

In spite of  the same value for the effective mass,  the dispersion curves at zero momentum have different slopes for motion parallel and normal to the field. The behavior of both curves is quite
different in both cases. 
This conclusion is to be expected since the magnetic field produces an
anisotropy in the system and the motion in these two directions have
different physical properties. However, the most notable result here is
the behavior of the effective mass $m_{eff}^{\nu}
\sim\mu_e eB/(M_W^2 - eB)$,
which increases without bound as $eB \to M_W^2$. For fields $eB \geq
M_W^2$, the neutrino magnetic mass problem, 
requires further
research along with the Higgs mechanism in external fields, taking into
account the results of refs. \cite{Olesen},\cite{Ambjorn}.

\section{Acknowledgments}
We thank A. Cabo, M. Ruiz-Altaba  and M. Torres for valuable discussions. 
The work of A.P.M, H.P.R and S.R.R. is partly supported by  
CONACYT,  A.P.M. thanks the South-South Fellowship Program of the Third
World Academy of Sciences for a  grant. H. P. R. thanks Professor
Virasoro, the ICTP High Energy Group, IAEA and UNESCO for hospitality
at the International Centre of Theoretical Physics.

\section{Appendix}
We present here the result of the calculation of $I_i$ far from the thresholds. Taking into account the
condition
\[
k_3^2-\omega ^2+d^2>\mu _e(k_3-\omega ),
\]

we get  \begin{eqnarray}
I_1=\frac{g^2r}{2(2\pi )^2}\frac{1}{k_3^2-\omega ^2+d^2}, \nonumber\\
I_2=\frac{g^2N_W}{2(2\pi )^2}\frac{1}{d(k_3^2-(\omega +d)^2)},  \nonumber \\
I_3=\frac{g^2N_W}{2(2\pi )^2}\frac{1}{k_3^2-(\omega +d)^2},  \nonumber \\
I_{4}=I_{1}\nonumber\\
I_5=0.
\end{eqnarray}
where $r=eB\mu_{e}$.

\newpage
\setlength{\unitlength}{0.240900pt}
\ifx\plotpoint\undefined\newsavebox{\plotpoint}\fi
\sbox{\plotpoint}{\rule[-0.500pt]{1.000pt}{1.000pt}}%
\begin{picture}(1500,900)(0,0)
\font\gnuplot=cmr10 at 10pt
\gnuplot
\sbox{\plotpoint}{\rule[-0.500pt]{1.000pt}{1.000pt}}%
\put(220.0,113.0){\rule[-0.500pt]{292.934pt}{1.000pt}}
\put(220.0,113.0){\rule[-0.500pt]{1.000pt}{184.048pt}}
\put(220.0,113.0){\rule[-0.500pt]{4.818pt}{1.000pt}}
\put(198,113){\makebox(0,0)[r]{0}}
\put(1416.0,113.0){\rule[-0.500pt]{4.818pt}{1.000pt}}
\put(220.0,266.0){\rule[-0.500pt]{4.818pt}{1.000pt}}
\put(198,266){\makebox(0,0)[r]{1}}
\put(1416.0,266.0){\rule[-0.500pt]{4.818pt}{1.000pt}}
\put(220.0,419.0){\rule[-0.500pt]{4.818pt}{1.000pt}}
\put(198,419){\makebox(0,0)[r]{2}}
\put(1416.0,419.0){\rule[-0.500pt]{4.818pt}{1.000pt}}
\put(220.0,571.0){\rule[-0.500pt]{4.818pt}{1.000pt}}
\put(198,571){\makebox(0,0)[r]{3}}
\put(1416.0,571.0){\rule[-0.500pt]{4.818pt}{1.000pt}}
\put(220.0,724.0){\rule[-0.500pt]{4.818pt}{1.000pt}}
\put(198,724){\makebox(0,0)[r]{4}}
\put(1416.0,724.0){\rule[-0.500pt]{4.818pt}{1.000pt}}
\put(220.0,877.0){\rule[-0.500pt]{4.818pt}{1.000pt}}
\put(198,877){\makebox(0,0)[r]{5}}
\put(1416.0,877.0){\rule[-0.500pt]{4.818pt}{1.000pt}}
\put(220.0,113.0){\rule[-0.500pt]{1.000pt}{4.818pt}}
\put(220,68){\makebox(0,0){0}}
\put(220.0,857.0){\rule[-0.500pt]{1.000pt}{4.818pt}}
\put(423.0,113.0){\rule[-0.500pt]{1.000pt}{4.818pt}}
\put(423,68){\makebox(0,0){1}}
\put(423.0,857.0){\rule[-0.500pt]{1.000pt}{4.818pt}}
\put(625.0,113.0){\rule[-0.500pt]{1.000pt}{4.818pt}}
\put(625,68){\makebox(0,0){2}}
\put(625.0,857.0){\rule[-0.500pt]{1.000pt}{4.818pt}}
\put(828.0,113.0){\rule[-0.500pt]{1.000pt}{4.818pt}}
\put(828,68){\makebox(0,0){3}}
\put(828.0,857.0){\rule[-0.500pt]{1.000pt}{4.818pt}}
\put(1031.0,113.0){\rule[-0.500pt]{1.000pt}{4.818pt}}
\put(1031,68){\makebox(0,0){4}}
\put(1031.0,857.0){\rule[-0.500pt]{1.000pt}{4.818pt}}
\put(1233.0,113.0){\rule[-0.500pt]{1.000pt}{4.818pt}}
\put(1233,68){\makebox(0,0){5}}
\put(1233.0,857.0){\rule[-0.500pt]{1.000pt}{4.818pt}}
\put(1436.0,113.0){\rule[-0.500pt]{1.000pt}{4.818pt}}
\put(1436,68){\makebox(0,0){6}}
\put(1436.0,857.0){\rule[-0.500pt]{1.000pt}{4.818pt}}
\put(220.0,113.0){\rule[-0.500pt]{292.934pt}{1.000pt}}
\put(1436.0,113.0){\rule[-0.500pt]{1.000pt}{184.048pt}}
\put(220.0,877.0){\rule[-0.500pt]{292.934pt}{1.000pt}}
\put(45,495){\makebox(0,0){${\omega}/{m_{e}}$}}
\put(828,23){\makebox(0,0){${k_{3}}/{m_{e}}$}}
\put(220.0,113.0){\rule[-0.500pt]{1.000pt}{184.048pt}}
\put(220,113){\usebox{\plotpoint}}
\multiput(220.00,114.86)(0.151,0.424){2}{\rule{1.650pt}{0.102pt}}
\multiput(220.00,110.92)(3.575,5.000){2}{\rule{0.825pt}{1.000pt}}
\put(234,124){\usebox{\plotpoint}}
\put(227,118){\usebox{\plotpoint}}
\put(220,260){\usebox{\plotpoint}}
\multiput(220.00,257.69)(0.932,-0.475){6}{\rule{2.250pt}{0.114pt}}
\multiput(220.00,257.92)(9.330,-7.000){2}{\rule{1.125pt}{1.000pt}}
\put(234,253){\usebox{\plotpoint}}
\put(234,253){\usebox{\plotpoint}}
\put(234,124){\usebox{\plotpoint}}
\put(242,129){\usebox{\plotpoint}}
\put(249,135){\usebox{\plotpoint}}
\put(242,129){\usebox{\plotpoint}}
\put(249,245){\usebox{\plotpoint}}
\put(234,253){\usebox{\plotpoint}}
\put(234,253){\usebox{\plotpoint}}
\put(234,253){\usebox{\plotpoint}}
\put(249,135){\usebox{\plotpoint}}
\put(256,140){\usebox{\plotpoint}}
\put(263,145){\usebox{\plotpoint}}
\put(256,140){\usebox{\plotpoint}}
\put(263,239){\usebox{\plotpoint}}
\put(257,242){\usebox{\plotpoint}}
\put(257,242){\usebox{\plotpoint}}
\put(249,245){\usebox{\plotpoint}}
\put(263,145){\usebox{\plotpoint}}
\put(270,151){\usebox{\plotpoint}}
\put(277,156){\usebox{\plotpoint}}
\put(270,151){\usebox{\plotpoint}}
\put(263,239){\usebox{\plotpoint}}
\put(271,236){\usebox{\plotpoint}}
\put(277,233){\usebox{\plotpoint}}
\put(271,236){\usebox{\plotpoint}}
\put(277,156){\usebox{\plotpoint}}
\put(284,161){\usebox{\plotpoint}}
\put(291,167){\usebox{\plotpoint}}
\put(284,161){\usebox{\plotpoint}}
\put(291,228){\usebox{\plotpoint}}
\put(284,231){\usebox{\plotpoint}}
\put(284,231){\usebox{\plotpoint}}
\put(277,233){\usebox{\plotpoint}}
\put(291,167){\usebox{\plotpoint}}
\put(298,172){\usebox{\plotpoint}}
\put(306,178){\usebox{\plotpoint}}
\put(298,172){\usebox{\plotpoint}}
\put(291,228){\usebox{\plotpoint}}
\put(299,226){\usebox{\plotpoint}}
\put(306,224){\usebox{\plotpoint}}
\put(299,226){\usebox{\plotpoint}}
\put(306,178){\usebox{\plotpoint}}
\put(312,183){\usebox{\plotpoint}}
\put(320,188){\usebox{\plotpoint}}
\put(312,183){\usebox{\plotpoint}}
\put(306,224){\usebox{\plotpoint}}
\put(319,221){\usebox{\plotpoint}}
\put(320,221){\usebox{\plotpoint}}
\put(319,221){\usebox{\plotpoint}}
\put(320,188){\usebox{\plotpoint}}
\put(326,194){\usebox{\plotpoint}}
\put(334,199){\usebox{\plotpoint}}
\put(326,194){\usebox{\plotpoint}}
\put(334,217){\usebox{\plotpoint}}
\put(324,221){\usebox{\plotpoint}}
\put(324,221){\usebox{\plotpoint}}
\put(320,221){\usebox{\plotpoint}}
\put(334,199){\usebox{\plotpoint}}
\put(339,206){\usebox{\plotpoint}}
\put(348,210){\usebox{\plotpoint}}
\put(339,206){\usebox{\plotpoint}}
\put(348,210){\usebox{\plotpoint}}
\put(334,217){\usebox{\plotpoint}}
\put(377,219){\usebox{\plotpoint}}
\put(363,221){\usebox{\plotpoint}}
\put(363,221){\usebox{\plotpoint}}
\put(375,222){\usebox{\plotpoint}}
\put(377,231){\usebox{\plotpoint}}
\put(375,222){\usebox{\plotpoint}}
\put(377,219){\usebox{\plotpoint}}
\put(378,220){\usebox{\plotpoint}}
\put(391,220){\usebox{\plotpoint}}
\put(378,220){\usebox{\plotpoint}}
\put(377,231){\usebox{\plotpoint}}
\put(387,235){\usebox{\plotpoint}}
\put(391,242){\usebox{\plotpoint}}
\put(387,235){\usebox{\plotpoint}}
\put(391,220){\usebox{\plotpoint}}
\put(392,220){\usebox{\plotpoint}}
\put(393,221){\usebox{\plotpoint}}
\put(392,220){\usebox{\plotpoint}}
\put(393,221){\usebox{\plotpoint}}
\put(400,225){\usebox{\plotpoint}}
\put(406,224){\usebox{\plotpoint}}
\put(400,225){\usebox{\plotpoint}}
\put(391,242){\usebox{\plotpoint}}
\put(400,246){\usebox{\plotpoint}}
\put(406,253){\usebox{\plotpoint}}
\put(400,246){\usebox{\plotpoint}}
\put(406,224){\usebox{\plotpoint}}
\put(411,227){\usebox{\plotpoint}}
\put(420,228){\usebox{\plotpoint}}
\put(411,227){\usebox{\plotpoint}}
\put(406,253){\usebox{\plotpoint}}
\put(414,257){\usebox{\plotpoint}}
\put(420,264){\usebox{\plotpoint}}
\put(414,257){\usebox{\plotpoint}}
\put(420,228){\usebox{\plotpoint}}
\put(423,229){\usebox{\plotpoint}}
\put(432,231){\usebox{\plotpoint}}
\put(423,229){\usebox{\plotpoint}}
\put(432,231){\usebox{\plotpoint}}
\put(433,232){\usebox{\plotpoint}}
\put(434,232){\usebox{\plotpoint}}
\put(433,232){\usebox{\plotpoint}}
\put(420,264){\usebox{\plotpoint}}
\put(428,268){\usebox{\plotpoint}}
\put(434,274){\usebox{\plotpoint}}
\put(428,268){\usebox{\plotpoint}}
\put(434,232){\usebox{\plotpoint}}
\put(441,237){\usebox{\plotpoint}}
\put(448,238){\usebox{\plotpoint}}
\put(441,237){\usebox{\plotpoint}}
\put(434,274){\usebox{\plotpoint}}
\put(443,279){\usebox{\plotpoint}}
\put(448,285){\usebox{\plotpoint}}
\put(443,279){\usebox{\plotpoint}}
\put(448,238){\usebox{\plotpoint}}
\put(451,240){\usebox{\plotpoint}}
\put(458,242){\usebox{\plotpoint}}
\put(451,240){\usebox{\plotpoint}}
\put(458,242){\usebox{\plotpoint}}
\put(461,244){\usebox{\plotpoint}}
\put(463,244){\usebox{\plotpoint}}
\put(461,244){\usebox{\plotpoint}}
\put(448,285){\usebox{\plotpoint}}
\put(457,290){\usebox{\plotpoint}}
\put(463,296){\usebox{\plotpoint}}
\put(457,290){\usebox{\plotpoint}}
\put(463,244){\usebox{\plotpoint}}
\put(469,248){\usebox{\plotpoint}}
\put(477,251){\usebox{\plotpoint}}
\put(469,248){\usebox{\plotpoint}}
\put(463,296){\usebox{\plotpoint}}
\put(471,300){\usebox{\plotpoint}}
\put(477,307){\usebox{\plotpoint}}
\put(471,300){\usebox{\plotpoint}}
\put(477,251){\usebox{\plotpoint}}
\put(478,252){\usebox{\plotpoint}}
\put(481,253){\usebox{\plotpoint}}
\put(478,252){\usebox{\plotpoint}}
\put(481,253){\usebox{\plotpoint}}
\put(486,257){\usebox{\plotpoint}}
\put(491,258){\usebox{\plotpoint}}
\put(486,257){\usebox{\plotpoint}}
\put(477,307){\usebox{\plotpoint}}
\put(485,311){\usebox{\plotpoint}}
\put(491,317){\usebox{\plotpoint}}
\put(485,311){\usebox{\plotpoint}}
\put(491,258){\usebox{\plotpoint}}
\put(495,261){\usebox{\plotpoint}}
\put(502,264){\usebox{\plotpoint}}
\put(495,261){\usebox{\plotpoint}}
\put(502,264){\usebox{\plotpoint}}
\put(504,265){\usebox{\plotpoint}}
\put(505,265){\usebox{\plotpoint}}
\put(504,265){\usebox{\plotpoint}}
\put(491,317){\usebox{\plotpoint}}
\put(499,322){\usebox{\plotpoint}}
\put(505,328){\usebox{\plotpoint}}
\put(499,322){\usebox{\plotpoint}}
\put(505,265){\usebox{\plotpoint}}
\put(511,270){\usebox{\plotpoint}}
\put(520,273){\usebox{\plotpoint}}
\put(511,270){\usebox{\plotpoint}}
\put(505,328){\usebox{\plotpoint}}
\put(514,333){\usebox{\plotpoint}}
\put(520,339){\usebox{\plotpoint}}
\put(514,333){\usebox{\plotpoint}}
\put(520,273){\usebox{\plotpoint}}
\put(520,274){\usebox{\plotpoint}}
\put(521,274){\usebox{\plotpoint}}
\put(520,274){\usebox{\plotpoint}}
\put(521,274){\usebox{\plotpoint}}
\put(527,279){\usebox{\plotpoint}}
\put(534,282){\usebox{\plotpoint}}
\put(527,279){\usebox{\plotpoint}}
\put(520,339){\usebox{\plotpoint}}
\put(528,344){\usebox{\plotpoint}}
\put(534,350){\usebox{\plotpoint}}
\put(528,344){\usebox{\plotpoint}}
\put(534,282){\usebox{\plotpoint}}
\put(536,284){\usebox{\plotpoint}}
\put(539,285){\usebox{\plotpoint}}
\put(536,284){\usebox{\plotpoint}}
\put(539,285){\usebox{\plotpoint}}
\put(544,289){\usebox{\plotpoint}}
\put(548,290){\usebox{\plotpoint}}
\put(544,289){\usebox{\plotpoint}}
\put(534,350){\usebox{\plotpoint}}
\put(542,354){\usebox{\plotpoint}}
\put(548,360){\usebox{\plotpoint}}
\put(542,354){\usebox{\plotpoint}}
\put(548,290){\usebox{\plotpoint}}
\put(552,293){\usebox{\plotpoint}}
\put(557,296){\usebox{\plotpoint}}
\put(552,293){\usebox{\plotpoint}}
\put(557,296){\usebox{\plotpoint}}
\put(560,298){\usebox{\plotpoint}}
\put(563,299){\usebox{\plotpoint}}
\put(560,298){\usebox{\plotpoint}}
\put(548,360){\usebox{\plotpoint}}
\put(556,365){\usebox{\plotpoint}}
\put(563,371){\usebox{\plotpoint}}
\put(556,365){\usebox{\plotpoint}}
\put(563,299){\usebox{\plotpoint}}
\put(567,303){\usebox{\plotpoint}}
\put(574,307){\usebox{\plotpoint}}
\put(567,303){\usebox{\plotpoint}}
\put(574,307){\usebox{\plotpoint}}
\put(576,308){\usebox{\plotpoint}}
\put(577,308){\usebox{\plotpoint}}
\put(576,308){\usebox{\plotpoint}}
\put(563,371){\usebox{\plotpoint}}
\put(571,376){\usebox{\plotpoint}}
\put(577,382){\usebox{\plotpoint}}
\put(571,376){\usebox{\plotpoint}}
\put(577,308){\usebox{\plotpoint}}
\put(583,313){\usebox{\plotpoint}}
\put(591,317){\usebox{\plotpoint}}
\put(583,313){\usebox{\plotpoint}}
\put(577,382){\usebox{\plotpoint}}
\put(585,387){\usebox{\plotpoint}}
\put(591,393){\usebox{\plotpoint}}
\put(585,387){\usebox{\plotpoint}}
\put(591,317){\usebox{\plotpoint}}
\put(591,317){\usebox{\plotpoint}}
\put(591,317){\usebox{\plotpoint}}
\put(591,317){\usebox{\plotpoint}}
\put(591,317){\usebox{\plotpoint}}
\put(598,323){\usebox{\plotpoint}}
\put(605,327){\usebox{\plotpoint}}
\put(598,323){\usebox{\plotpoint}}
\put(591,393){\usebox{\plotpoint}}
\put(599,397){\usebox{\plotpoint}}
\put(605,404){\usebox{\plotpoint}}
\put(599,397){\usebox{\plotpoint}}
\put(605,327){\usebox{\plotpoint}}
\put(606,327){\usebox{\plotpoint}}
\put(608,328){\usebox{\plotpoint}}
\put(606,327){\usebox{\plotpoint}}
\put(608,328){\usebox{\plotpoint}}
\put(613,333){\usebox{\plotpoint}}
\put(620,336){\usebox{\plotpoint}}
\put(613,333){\usebox{\plotpoint}}
\put(605,404){\usebox{\plotpoint}}
\put(613,408){\usebox{\plotpoint}}
\put(620,414){\usebox{\plotpoint}}
\put(613,408){\usebox{\plotpoint}}
\put(620,336){\usebox{\plotpoint}}
\put(621,338){\usebox{\plotpoint}}
\put(624,339){\usebox{\plotpoint}}
\put(621,338){\usebox{\plotpoint}}
\put(624,339){\usebox{\plotpoint}}
\put(628,343){\usebox{\plotpoint}}
\put(634,346){\usebox{\plotpoint}}
\put(628,343){\usebox{\plotpoint}}
\put(620,414){\usebox{\plotpoint}}
\put(628,419){\usebox{\plotpoint}}
\put(634,425){\usebox{\plotpoint}}
\put(628,419){\usebox{\plotpoint}}
\put(634,346){\usebox{\plotpoint}}
\put(636,348){\usebox{\plotpoint}}
\put(640,350){\usebox{\plotpoint}}
\put(636,348){\usebox{\plotpoint}}
\put(640,350){\usebox{\plotpoint}}
\put(644,353){\usebox{\plotpoint}}
\put(648,355){\usebox{\plotpoint}}
\put(644,353){\usebox{\plotpoint}}
\put(634,425){\usebox{\plotpoint}}
\put(642,430){\usebox{\plotpoint}}
\put(648,436){\usebox{\plotpoint}}
\put(642,430){\usebox{\plotpoint}}
\put(648,355){\usebox{\plotpoint}}
\put(651,358){\usebox{\plotpoint}}
\put(655,360){\usebox{\plotpoint}}
\put(651,358){\usebox{\plotpoint}}
\put(655,360){\usebox{\plotpoint}}
\put(659,363){\usebox{\plotpoint}}
\put(662,365){\usebox{\plotpoint}}
\put(659,363){\usebox{\plotpoint}}
\put(648,436){\usebox{\plotpoint}}
\put(656,441){\usebox{\plotpoint}}
\put(662,447){\usebox{\plotpoint}}
\put(656,441){\usebox{\plotpoint}}
\put(662,365){\usebox{\plotpoint}}
\put(666,369){\usebox{\plotpoint}}
\put(671,371){\usebox{\plotpoint}}
\put(666,369){\usebox{\plotpoint}}
\put(671,371){\usebox{\plotpoint}}
\put(674,374){\usebox{\plotpoint}}
\put(677,375){\usebox{\plotpoint}}
\put(674,374){\usebox{\plotpoint}}
\put(662,447){\usebox{\plotpoint}}
\put(670,451){\usebox{\plotpoint}}
\put(677,457){\usebox{\plotpoint}}
\put(670,451){\usebox{\plotpoint}}
\put(677,375){\usebox{\plotpoint}}
\put(681,379){\usebox{\plotpoint}}
\put(687,382){\usebox{\plotpoint}}
\put(681,379){\usebox{\plotpoint}}
\put(687,382){\usebox{\plotpoint}}
\put(689,384){\usebox{\plotpoint}}
\put(691,385){\usebox{\plotpoint}}
\put(689,384){\usebox{\plotpoint}}
\put(677,457){\usebox{\plotpoint}}
\put(685,462){\usebox{\plotpoint}}
\put(691,468){\usebox{\plotpoint}}
\put(685,462){\usebox{\plotpoint}}
\put(691,385){\usebox{\plotpoint}}
\put(696,389){\usebox{\plotpoint}}
\put(702,393){\usebox{\plotpoint}}
\put(696,389){\usebox{\plotpoint}}
\put(702,393){\usebox{\plotpoint}}
\put(704,394){\usebox{\plotpoint}}
\put(705,395){\usebox{\plotpoint}}
\put(704,394){\usebox{\plotpoint}}
\put(691,468){\usebox{\plotpoint}}
\put(699,473){\usebox{\plotpoint}}
\put(705,479){\usebox{\plotpoint}}
\put(699,473){\usebox{\plotpoint}}
\put(705,395){\usebox{\plotpoint}}
\put(710,400){\usebox{\plotpoint}}
\put(718,404){\usebox{\plotpoint}}
\put(710,400){\usebox{\plotpoint}}
\put(718,404){\usebox{\plotpoint}}
\put(718,404){\usebox{\plotpoint}}
\put(720,405){\usebox{\plotpoint}}
\put(718,404){\usebox{\plotpoint}}
\put(705,479){\usebox{\plotpoint}}
\put(713,484){\usebox{\plotpoint}}
\put(720,490){\usebox{\plotpoint}}
\put(713,484){\usebox{\plotpoint}}
\put(720,405){\usebox{\plotpoint}}
\put(725,410){\usebox{\plotpoint}}
\put(733,414){\usebox{\plotpoint}}
\put(725,410){\usebox{\plotpoint}}
\put(733,414){\usebox{\plotpoint}}
\put(733,415){\usebox{\plotpoint}}
\put(734,415){\usebox{\plotpoint}}
\put(733,415){\usebox{\plotpoint}}
\put(720,490){\usebox{\plotpoint}}
\put(728,494){\usebox{\plotpoint}}
\put(734,500){\usebox{\plotpoint}}
\put(728,494){\usebox{\plotpoint}}
\put(734,415){\usebox{\plotpoint}}
\put(740,420){\usebox{\plotpoint}}
\put(748,425){\usebox{\plotpoint}}
\put(740,420){\usebox{\plotpoint}}
\put(734,500){\usebox{\plotpoint}}
\put(742,505){\usebox{\plotpoint}}
\put(748,511){\usebox{\plotpoint}}
\put(742,505){\usebox{\plotpoint}}
\put(748,425){\usebox{\plotpoint}}
\put(748,425){\usebox{\plotpoint}}
\put(748,425){\usebox{\plotpoint}}
\put(748,425){\usebox{\plotpoint}}
\put(748,425){\usebox{\plotpoint}}
\put(755,431){\usebox{\plotpoint}}
\put(762,435){\usebox{\plotpoint}}
\put(755,431){\usebox{\plotpoint}}
\put(748,511){\usebox{\plotpoint}}
\put(756,516){\usebox{\plotpoint}}
\put(762,522){\usebox{\plotpoint}}
\put(756,516){\usebox{\plotpoint}}
\put(762,435){\usebox{\plotpoint}}
\put(763,435){\usebox{\plotpoint}}
\put(763,436){\usebox{\plotpoint}}
\put(763,435){\usebox{\plotpoint}}
\put(763,436){\usebox{\plotpoint}}
\put(769,441){\usebox{\plotpoint}}
\put(777,445){\usebox{\plotpoint}}
\put(769,441){\usebox{\plotpoint}}
\put(762,522){\usebox{\plotpoint}}
\put(770,527){\usebox{\plotpoint}}
\put(777,533){\usebox{\plotpoint}}
\put(770,527){\usebox{\plotpoint}}
\put(777,445){\usebox{\plotpoint}}
\put(777,446){\usebox{\plotpoint}}
\put(778,447){\usebox{\plotpoint}}
\put(777,446){\usebox{\plotpoint}}
\put(778,447){\usebox{\plotpoint}}
\put(784,452){\usebox{\plotpoint}}
\put(791,456){\usebox{\plotpoint}}
\put(784,452){\usebox{\plotpoint}}
\put(777,533){\usebox{\plotpoint}}
\put(785,537){\usebox{\plotpoint}}
\put(791,543){\usebox{\plotpoint}}
\put(785,537){\usebox{\plotpoint}}
\put(791,456){\usebox{\plotpoint}}
\put(792,457){\usebox{\plotpoint}}
\put(793,457){\usebox{\plotpoint}}
\put(792,457){\usebox{\plotpoint}}
\put(793,457){\usebox{\plotpoint}}
\put(799,462){\usebox{\plotpoint}}
\put(805,466){\usebox{\plotpoint}}
\put(799,462){\usebox{\plotpoint}}
\put(791,543){\usebox{\plotpoint}}
\put(799,548){\usebox{\plotpoint}}
\put(805,554){\usebox{\plotpoint}}
\put(799,548){\usebox{\plotpoint}}
\put(805,466){\usebox{\plotpoint}}
\put(806,467){\usebox{\plotpoint}}
\put(808,468){\usebox{\plotpoint}}
\put(806,467){\usebox{\plotpoint}}
\put(808,468){\usebox{\plotpoint}}
\put(813,473){\usebox{\plotpoint}}
\put(819,476){\usebox{\plotpoint}}
\put(813,473){\usebox{\plotpoint}}
\put(805,554){\usebox{\plotpoint}}
\put(813,559){\usebox{\plotpoint}}
\put(819,565){\usebox{\plotpoint}}
\put(813,559){\usebox{\plotpoint}}
\put(819,476){\usebox{\plotpoint}}
\put(821,478){\usebox{\plotpoint}}
\put(823,479){\usebox{\plotpoint}}
\put(821,478){\usebox{\plotpoint}}
\put(823,479){\usebox{\plotpoint}}
\put(828,483){\usebox{\plotpoint}}
\put(834,487){\usebox{\plotpoint}}
\put(828,483){\usebox{\plotpoint}}
\put(819,565){\usebox{\plotpoint}}
\put(827,570){\usebox{\plotpoint}}
\put(834,576){\usebox{\plotpoint}}
\put(827,570){\usebox{\plotpoint}}
\put(834,487){\usebox{\plotpoint}}
\put(836,488){\usebox{\plotpoint}}
\put(838,490){\usebox{\plotpoint}}
\put(836,488){\usebox{\plotpoint}}
\put(838,490){\usebox{\plotpoint}}
\put(842,494){\usebox{\plotpoint}}
\put(848,497){\usebox{\plotpoint}}
\put(842,494){\usebox{\plotpoint}}
\put(834,576){\usebox{\plotpoint}}
\put(842,580){\usebox{\plotpoint}}
\put(848,586){\usebox{\plotpoint}}
\put(842,580){\usebox{\plotpoint}}
\put(848,497){\usebox{\plotpoint}}
\put(850,499){\usebox{\plotpoint}}
\put(853,500){\usebox{\plotpoint}}
\put(850,499){\usebox{\plotpoint}}
\put(853,500){\usebox{\plotpoint}}
\put(857,504){\usebox{\plotpoint}}
\put(862,507){\usebox{\plotpoint}}
\put(857,504){\usebox{\plotpoint}}
\put(848,586){\usebox{\plotpoint}}
\put(856,591){\usebox{\plotpoint}}
\put(862,597){\usebox{\plotpoint}}
\put(856,591){\usebox{\plotpoint}}
\put(862,507){\usebox{\plotpoint}}
\put(865,509){\usebox{\plotpoint}}
\put(867,511){\usebox{\plotpoint}}
\put(865,509){\usebox{\plotpoint}}
\put(867,511){\usebox{\plotpoint}}
\put(871,515){\usebox{\plotpoint}}
\put(877,518){\usebox{\plotpoint}}
\put(871,515){\usebox{\plotpoint}}
\put(862,597){\usebox{\plotpoint}}
\put(870,602){\usebox{\plotpoint}}
\put(877,608){\usebox{\plotpoint}}
\put(870,602){\usebox{\plotpoint}}
\put(877,518){\usebox{\plotpoint}}
\put(879,520){\usebox{\plotpoint}}
\put(882,522){\usebox{\plotpoint}}
\put(879,520){\usebox{\plotpoint}}
\put(882,522){\usebox{\plotpoint}}
\put(886,526){\usebox{\plotpoint}}
\put(891,528){\usebox{\plotpoint}}
\put(886,526){\usebox{\plotpoint}}
\put(877,608){\usebox{\plotpoint}}
\put(884,613){\usebox{\plotpoint}}
\put(891,619){\usebox{\plotpoint}}
\put(884,613){\usebox{\plotpoint}}
\put(891,528){\usebox{\plotpoint}}
\put(894,531){\usebox{\plotpoint}}
\put(897,533){\usebox{\plotpoint}}
\put(894,531){\usebox{\plotpoint}}
\put(897,533){\usebox{\plotpoint}}
\put(901,536){\usebox{\plotpoint}}
\put(905,539){\usebox{\plotpoint}}
\put(901,536){\usebox{\plotpoint}}
\put(891,619){\usebox{\plotpoint}}
\put(899,623){\usebox{\plotpoint}}
\put(905,630){\usebox{\plotpoint}}
\put(899,623){\usebox{\plotpoint}}
\put(905,539){\usebox{\plotpoint}}
\put(908,541){\usebox{\plotpoint}}
\put(912,543){\usebox{\plotpoint}}
\put(908,541){\usebox{\plotpoint}}
\put(912,543){\usebox{\plotpoint}}
\put(915,547){\usebox{\plotpoint}}
\put(919,549){\usebox{\plotpoint}}
\put(915,547){\usebox{\plotpoint}}
\put(905,630){\usebox{\plotpoint}}
\put(913,634){\usebox{\plotpoint}}
\put(919,640){\usebox{\plotpoint}}
\put(913,634){\usebox{\plotpoint}}
\put(919,549){\usebox{\plotpoint}}
\put(922,552){\usebox{\plotpoint}}
\put(926,554){\usebox{\plotpoint}}
\put(922,552){\usebox{\plotpoint}}
\put(926,554){\usebox{\plotpoint}}
\put(930,557){\usebox{\plotpoint}}
\put(934,559){\usebox{\plotpoint}}
\put(930,557){\usebox{\plotpoint}}
\put(919,640){\usebox{\plotpoint}}
\put(927,645){\usebox{\plotpoint}}
\put(934,651){\usebox{\plotpoint}}
\put(927,645){\usebox{\plotpoint}}
\put(934,559){\usebox{\plotpoint}}
\put(937,562){\usebox{\plotpoint}}
\put(941,565){\usebox{\plotpoint}}
\put(937,562){\usebox{\plotpoint}}
\put(941,565){\usebox{\plotpoint}}
\put(944,568){\usebox{\plotpoint}}
\put(948,570){\usebox{\plotpoint}}
\put(944,568){\usebox{\plotpoint}}
\put(934,651){\usebox{\plotpoint}}
\put(942,656){\usebox{\plotpoint}}
\put(948,662){\usebox{\plotpoint}}
\put(942,656){\usebox{\plotpoint}}
\put(948,570){\usebox{\plotpoint}}
\put(951,573){\usebox{\plotpoint}}
\put(956,576){\usebox{\plotpoint}}
\put(951,573){\usebox{\plotpoint}}
\put(956,576){\usebox{\plotpoint}}
\put(959,578){\usebox{\plotpoint}}
\put(962,580){\usebox{\plotpoint}}
\put(959,578){\usebox{\plotpoint}}
\put(948,662){\usebox{\plotpoint}}
\put(956,667){\usebox{\plotpoint}}
\put(962,673){\usebox{\plotpoint}}
\put(956,667){\usebox{\plotpoint}}
\put(962,580){\usebox{\plotpoint}}
\put(966,584){\usebox{\plotpoint}}
\put(970,586){\usebox{\plotpoint}}
\put(966,584){\usebox{\plotpoint}}
\put(970,586){\usebox{\plotpoint}}
\put(973,589){\usebox{\plotpoint}}
\put(976,591){\usebox{\plotpoint}}
\put(973,589){\usebox{\plotpoint}}
\put(962,673){\usebox{\plotpoint}}
\put(970,677){\usebox{\plotpoint}}
\put(976,683){\usebox{\plotpoint}}
\put(970,677){\usebox{\plotpoint}}
\put(976,591){\usebox{\plotpoint}}
\put(980,594){\usebox{\plotpoint}}
\put(985,597){\usebox{\plotpoint}}
\put(980,594){\usebox{\plotpoint}}
\put(985,597){\usebox{\plotpoint}}
\put(987,600){\usebox{\plotpoint}}
\put(991,601){\usebox{\plotpoint}}
\put(987,600){\usebox{\plotpoint}}
\put(976,683){\usebox{\plotpoint}}
\put(984,688){\usebox{\plotpoint}}
\put(991,694){\usebox{\plotpoint}}
\put(984,688){\usebox{\plotpoint}}
\put(991,601){\usebox{\plotpoint}}
\put(995,605){\usebox{\plotpoint}}
\put(1000,608){\usebox{\plotpoint}}
\put(995,605){\usebox{\plotpoint}}
\put(1000,608){\usebox{\plotpoint}}
\put(1002,610){\usebox{\plotpoint}}
\put(1005,612){\usebox{\plotpoint}}
\put(1002,610){\usebox{\plotpoint}}
\put(991,694){\usebox{\plotpoint}}
\put(999,699){\usebox{\plotpoint}}
\put(1005,705){\usebox{\plotpoint}}
\put(999,699){\usebox{\plotpoint}}
\put(1005,612){\usebox{\plotpoint}}
\put(1009,616){\usebox{\plotpoint}}
\put(1014,619){\usebox{\plotpoint}}
\put(1009,616){\usebox{\plotpoint}}
\put(1014,619){\usebox{\plotpoint}}
\put(1016,621){\usebox{\plotpoint}}
\put(1019,622){\usebox{\plotpoint}}
\put(1016,621){\usebox{\plotpoint}}
\put(1005,705){\usebox{\plotpoint}}
\put(1013,710){\usebox{\plotpoint}}
\put(1019,716){\usebox{\plotpoint}}
\put(1013,710){\usebox{\plotpoint}}
\put(1019,622){\usebox{\plotpoint}}
\put(1023,626){\usebox{\plotpoint}}
\put(1029,630){\usebox{\plotpoint}}
\put(1023,626){\usebox{\plotpoint}}
\put(1029,630){\usebox{\plotpoint}}
\put(1031,632){\usebox{\plotpoint}}
\put(1034,633){\usebox{\plotpoint}}
\put(1031,632){\usebox{\plotpoint}}
\put(1019,716){\usebox{\plotpoint}}
\put(1027,720){\usebox{\plotpoint}}
\put(1034,726){\usebox{\plotpoint}}
\put(1027,720){\usebox{\plotpoint}}
\put(1034,633){\usebox{\plotpoint}}
\put(1038,637){\usebox{\plotpoint}}
\put(1043,640){\usebox{\plotpoint}}
\put(1038,637){\usebox{\plotpoint}}
\put(1043,640){\usebox{\plotpoint}}
\put(1045,642){\usebox{\plotpoint}}
\put(1048,643){\usebox{\plotpoint}}
\put(1045,642){\usebox{\plotpoint}}
\put(1034,726){\usebox{\plotpoint}}
\put(1041,731){\usebox{\plotpoint}}
\put(1048,737){\usebox{\plotpoint}}
\put(1041,731){\usebox{\plotpoint}}
\put(1048,643){\usebox{\plotpoint}}
\put(1052,648){\usebox{\plotpoint}}
\put(1058,651){\usebox{\plotpoint}}
\put(1052,648){\usebox{\plotpoint}}
\put(1058,651){\usebox{\plotpoint}}
\put(1060,653){\usebox{\plotpoint}}
\put(1062,654){\usebox{\plotpoint}}
\put(1060,653){\usebox{\plotpoint}}
\put(1048,737){\usebox{\plotpoint}}
\put(1056,742){\usebox{\plotpoint}}
\put(1062,748){\usebox{\plotpoint}}
\put(1056,742){\usebox{\plotpoint}}
\put(1062,654){\usebox{\plotpoint}}
\put(1067,658){\usebox{\plotpoint}}
\put(1072,662){\usebox{\plotpoint}}
\put(1067,658){\usebox{\plotpoint}}
\put(1072,662){\usebox{\plotpoint}}
\put(1074,663){\usebox{\plotpoint}}
\put(1076,665){\usebox{\plotpoint}}
\put(1074,663){\usebox{\plotpoint}}
\put(1062,748){\usebox{\plotpoint}}
\put(1070,753){\usebox{\plotpoint}}
\put(1076,759){\usebox{\plotpoint}}
\put(1070,753){\usebox{\plotpoint}}
\put(1076,665){\usebox{\plotpoint}}
\put(1081,669){\usebox{\plotpoint}}
\put(1087,673){\usebox{\plotpoint}}
\put(1081,669){\usebox{\plotpoint}}
\put(1087,673){\usebox{\plotpoint}}
\put(1089,674){\usebox{\plotpoint}}
\put(1091,675){\usebox{\plotpoint}}
\put(1089,674){\usebox{\plotpoint}}
\put(1076,759){\usebox{\plotpoint}}
\put(1084,763){\usebox{\plotpoint}}
\put(1091,769){\usebox{\plotpoint}}
\put(1084,763){\usebox{\plotpoint}}
\put(1091,675){\usebox{\plotpoint}}
\put(1095,680){\usebox{\plotpoint}}
\put(1102,683){\usebox{\plotpoint}}
\put(1095,680){\usebox{\plotpoint}}
\put(1102,683){\usebox{\plotpoint}}
\put(1103,685){\usebox{\plotpoint}}
\put(1105,686){\usebox{\plotpoint}}
\put(1103,685){\usebox{\plotpoint}}
\put(1091,769){\usebox{\plotpoint}}
\put(1099,774){\usebox{\plotpoint}}
\put(1105,780){\usebox{\plotpoint}}
\put(1099,774){\usebox{\plotpoint}}
\put(1105,686){\usebox{\plotpoint}}
\put(1110,690){\usebox{\plotpoint}}
\put(1116,694){\usebox{\plotpoint}}
\put(1110,690){\usebox{\plotpoint}}
\put(1116,694){\usebox{\plotpoint}}
\put(1117,695){\usebox{\plotpoint}}
\put(1119,696){\usebox{\plotpoint}}
\put(1117,695){\usebox{\plotpoint}}
\put(1105,780){\usebox{\plotpoint}}
\put(1113,785){\usebox{\plotpoint}}
\put(1119,791){\usebox{\plotpoint}}
\put(1113,785){\usebox{\plotpoint}}
\put(1119,696){\usebox{\plotpoint}}
\put(1124,701){\usebox{\plotpoint}}
\put(1131,705){\usebox{\plotpoint}}
\put(1124,701){\usebox{\plotpoint}}
\put(1131,705){\usebox{\plotpoint}}
\put(1132,706){\usebox{\plotpoint}}
\put(1133,707){\usebox{\plotpoint}}
\put(1132,706){\usebox{\plotpoint}}
\put(1119,791){\usebox{\plotpoint}}
\put(1127,796){\usebox{\plotpoint}}
\put(1133,802){\usebox{\plotpoint}}
\put(1127,796){\usebox{\plotpoint}}
\put(1133,707){\usebox{\plotpoint}}
\put(1139,712){\usebox{\plotpoint}}
\put(1145,716){\usebox{\plotpoint}}
\put(1139,712){\usebox{\plotpoint}}
\put(1145,716){\usebox{\plotpoint}}
\put(1146,717){\usebox{\plotpoint}}
\put(1148,717){\usebox{\plotpoint}}
\put(1146,717){\usebox{\plotpoint}}
\put(1133,802){\usebox{\plotpoint}}
\put(1141,806){\usebox{\plotpoint}}
\put(1148,812){\usebox{\plotpoint}}
\put(1141,806){\usebox{\plotpoint}}
\put(1148,717){\usebox{\plotpoint}}
\put(1153,722){\usebox{\plotpoint}}
\put(1160,726){\usebox{\plotpoint}}
\put(1153,722){\usebox{\plotpoint}}
\put(1160,726){\usebox{\plotpoint}}
\put(1161,727){\usebox{\plotpoint}}
\put(1162,728){\usebox{\plotpoint}}
\put(1161,727){\usebox{\plotpoint}}
\put(1148,812){\usebox{\plotpoint}}
\put(1156,817){\usebox{\plotpoint}}
\put(1162,823){\usebox{\plotpoint}}
\put(1156,817){\usebox{\plotpoint}}
\put(1162,728){\usebox{\plotpoint}}
\put(1167,733){\usebox{\plotpoint}}
\put(1174,737){\usebox{\plotpoint}}
\put(1167,733){\usebox{\plotpoint}}
\put(1174,737){\usebox{\plotpoint}}
\put(1175,738){\usebox{\plotpoint}}
\put(1176,739){\usebox{\plotpoint}}
\put(1175,738){\usebox{\plotpoint}}
\put(1162,823){\usebox{\plotpoint}}
\put(1170,828){\usebox{\plotpoint}}
\put(1176,834){\usebox{\plotpoint}}
\put(1170,828){\usebox{\plotpoint}}
\put(1176,739){\usebox{\plotpoint}}
\put(1182,744){\usebox{\plotpoint}}
\put(1189,748){\usebox{\plotpoint}}
\put(1182,744){\usebox{\plotpoint}}
\put(1189,748){\usebox{\plotpoint}}
\put(1189,749){\usebox{\plotpoint}}
\put(1191,749){\usebox{\plotpoint}}
\put(1189,749){\usebox{\plotpoint}}
\put(1176,834){\usebox{\plotpoint}}
\put(1184,839){\usebox{\plotpoint}}
\put(1191,845){\usebox{\plotpoint}}
\put(1184,839){\usebox{\plotpoint}}
\put(1191,749){\usebox{\plotpoint}}
\put(1196,754){\usebox{\plotpoint}}
\put(1203,759){\usebox{\plotpoint}}
\put(1196,754){\usebox{\plotpoint}}
\put(1203,759){\usebox{\plotpoint}}
\put(1204,759){\usebox{\plotpoint}}
\put(1205,760){\usebox{\plotpoint}}
\put(1204,759){\usebox{\plotpoint}}
\put(1191,845){\usebox{\plotpoint}}
\put(1198,849){\usebox{\plotpoint}}
\put(1205,855){\usebox{\plotpoint}}
\put(1198,849){\usebox{\plotpoint}}
\put(1205,760){\usebox{\plotpoint}}
\put(1210,765){\usebox{\plotpoint}}
\put(1218,769){\usebox{\plotpoint}}
\put(1210,765){\usebox{\plotpoint}}
\put(1218,769){\usebox{\plotpoint}}
\put(1218,770){\usebox{\plotpoint}}
\put(1219,770){\usebox{\plotpoint}}
\put(1218,770){\usebox{\plotpoint}}
\put(1205,855){\usebox{\plotpoint}}
\put(1213,860){\usebox{\plotpoint}}
\put(1219,866){\usebox{\plotpoint}}
\put(1213,860){\usebox{\plotpoint}}
\put(1219,770){\usebox{\plotpoint}}
\put(1225,776){\usebox{\plotpoint}}
\put(1232,780){\usebox{\plotpoint}}
\put(1225,776){\usebox{\plotpoint}}
\put(1232,780){\usebox{\plotpoint}}
\put(1233,781){\usebox{\plotpoint}}
\put(1233,781){\usebox{\plotpoint}}
\put(1233,781){\usebox{\plotpoint}}
\put(1219,866){\usebox{\plotpoint}}
\put(1227,871){\usebox{\plotpoint}}
\put(1233,877){\usebox{\plotpoint}}
\put(1227,871){\usebox{\plotpoint}}
\end{picture}
\begin{center}
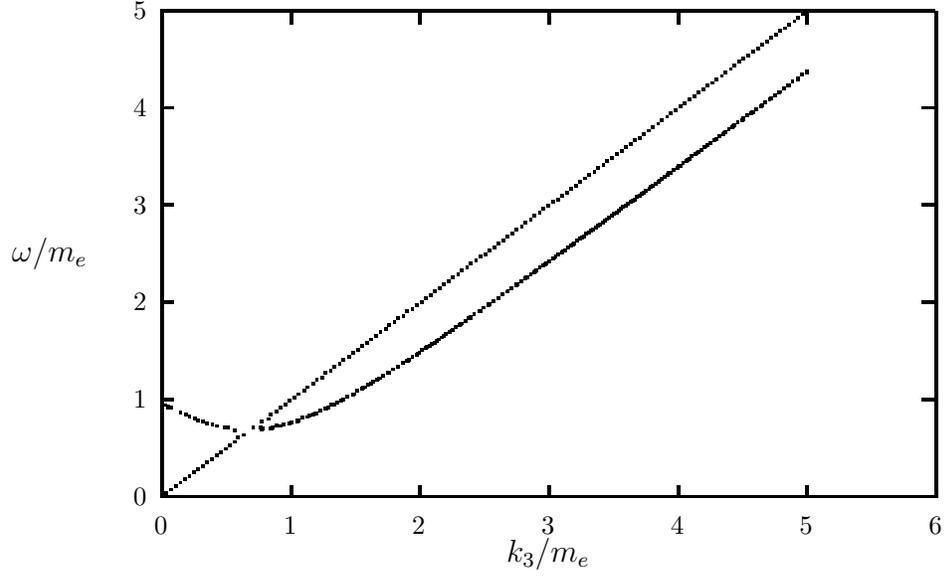
{\it Fig 1. Curve of propagation of neutrino paralell to the magnetic field}
\end{center}
\vspace{1cm}

\begin{picture}(1500,900)(0,0)
\font\gnuplot=cmr10 at 10pt
\gnuplot
\put(220.0,113.0){\rule[-0.500pt]{292.934pt}{1.000pt}}
\put(220.0,113.0){\rule[-0.500pt]{1.000pt}{184.048pt}}
\put(220.0,113.0){\rule[-0.500pt]{4.818pt}{1.000pt}}
\put(198,113){\makebox(0,0)[r]{0}}
\put(1416.0,113.0){\rule[-0.500pt]{4.818pt}{1.000pt}}
\put(220.0,266.0){\rule[-0.500pt]{4.818pt}{1.000pt}}
\put(198,266){\makebox(0,0)[r]{1}}
\put(1416.0,266.0){\rule[-0.500pt]{4.818pt}{1.000pt}}
\put(220.0,419.0){\rule[-0.500pt]{4.818pt}{1.000pt}}
\put(198,419){\makebox(0,0)[r]{2}}
\put(1416.0,419.0){\rule[-0.500pt]{4.818pt}{1.000pt}}
\put(220.0,571.0){\rule[-0.500pt]{4.818pt}{1.000pt}}
\put(198,571){\makebox(0,0)[r]{3}}
\put(1416.0,571.0){\rule[-0.500pt]{4.818pt}{1.000pt}}
\put(220.0,724.0){\rule[-0.500pt]{4.818pt}{1.000pt}}
\put(198,724){\makebox(0,0)[r]{4}}
\put(1416.0,724.0){\rule[-0.500pt]{4.818pt}{1.000pt}}
\put(220.0,877.0){\rule[-0.500pt]{4.818pt}{1.000pt}}
\put(198,877){\makebox(0,0)[r]{5}}
\put(1416.0,877.0){\rule[-0.500pt]{4.818pt}{1.000pt}}
\put(220.0,113.0){\rule[-0.500pt]{1.000pt}{4.818pt}}
\put(220,68){\makebox(0,0){0}}
\put(220.0,857.0){\rule[-0.500pt]{1.000pt}{4.818pt}}
\put(463.0,113.0){\rule[-0.500pt]{1.000pt}{4.818pt}}
\put(463,68){\makebox(0,0){1}}
\put(463.0,857.0){\rule[-0.500pt]{1.000pt}{4.818pt}}
\put(706.0,113.0){\rule[-0.500pt]{1.000pt}{4.818pt}}
\put(706,68){\makebox(0,0){2}}
\put(706.0,857.0){\rule[-0.500pt]{1.000pt}{4.818pt}}
\put(950.0,113.0){\rule[-0.500pt]{1.000pt}{4.818pt}}
\put(950,68){\makebox(0,0){3}}
\put(950.0,857.0){\rule[-0.500pt]{1.000pt}{4.818pt}}
\put(1193.0,113.0){\rule[-0.500pt]{1.000pt}{4.818pt}}
\put(1193,68){\makebox(0,0){4}}
\put(1193.0,857.0){\rule[-0.500pt]{1.000pt}{4.818pt}}
\put(1436.0,113.0){\rule[-0.500pt]{1.000pt}{4.818pt}}
\put(1436,68){\makebox(0,0){5}}
\put(1436.0,857.0){\rule[-0.500pt]{1.000pt}{4.818pt}}
\put(220.0,113.0){\rule[-0.500pt]{292.934pt}{1.000pt}}
\put(1436.0,113.0){\rule[-0.500pt]{1.000pt}{184.048pt}}
\put(220.0,877.0){\rule[-0.500pt]{292.934pt}{1.000pt}}
\put(45,495){\makebox(0,0){$ {\omega}/{m_{e}}$}}
\put(828,23){\makebox(0,0){$ {k_{\perp}}/{m_{e}}$}}
\put(220.0,113.0){\rule[-0.500pt]{1.000pt}{184.048pt}}
\put(220,113){\usebox{\plotpoint}}
\put(220,111.42){\rule{8.913pt}{1.000pt}}
\multiput(220.00,110.92)(18.500,1.000){2}{\rule{4.457pt}{1.000pt}}
\put(258,114){\usebox{\plotpoint}}
\put(257,114){\usebox{\plotpoint}}
\put(220,259){\usebox{\plotpoint}}
\put(220,257.92){\rule{7.227pt}{1.000pt}}
\multiput(220.00,256.92)(15.000,2.000){2}{\rule{3.613pt}{1.000pt}}
\put(258,262){\usebox{\plotpoint}}
\put(250,261){\usebox{\plotpoint}}
\put(258,114){\usebox{\plotpoint}}
\put(285,120){\usebox{\plotpoint}}
\put(296,121){\usebox{\plotpoint}}
\put(285,120){\usebox{\plotpoint}}
\put(258,262){\usebox{\plotpoint}}
\put(279,267){\usebox{\plotpoint}}
\put(296,273){\usebox{\plotpoint}}
\put(279,267){\usebox{\plotpoint}}
\put(296,121){\usebox{\plotpoint}}
\put(306,131){\usebox{\plotpoint}}
\put(316,137){\usebox{\plotpoint}}
\put(306,131){\usebox{\plotpoint}}
\put(316,137){\usebox{\plotpoint}}
\put(328,140){\usebox{\plotpoint}}
\put(334,141){\usebox{\plotpoint}}
\put(328,140){\usebox{\plotpoint}}
\put(296,273){\usebox{\plotpoint}}
\put(304,275){\usebox{\plotpoint}}
\put(310,280){\usebox{\plotpoint}}
\put(304,275){\usebox{\plotpoint}}
\put(310,280){\usebox{\plotpoint}}
\put(327,285){\usebox{\plotpoint}}
\put(334,288){\usebox{\plotpoint}}
\put(327,285){\usebox{\plotpoint}}
\put(334,141){\usebox{\plotpoint}}
\put(349,151){\usebox{\plotpoint}}
\put(372,160){\usebox{\plotpoint}}
\put(349,151){\usebox{\plotpoint}}
\put(334,288){\usebox{\plotpoint}}
\put(350,294){\usebox{\plotpoint}}
\put(365,304){\usebox{\plotpoint}}
\put(350,294){\usebox{\plotpoint}}
\put(365,304){\usebox{\plotpoint}}
\put(370,305){\usebox{\plotpoint}}
\put(372,306){\usebox{\plotpoint}}
\put(370,305){\usebox{\plotpoint}}
\put(372,160){\usebox{\plotpoint}}
\put(373,160){\usebox{\plotpoint}}
\put(373,161){\usebox{\plotpoint}}
\put(373,160){\usebox{\plotpoint}}
\put(373,161){\usebox{\plotpoint}}
\put(391,173){\usebox{\plotpoint}}
\put(410,179){\usebox{\plotpoint}}
\put(391,173){\usebox{\plotpoint}}
\put(372,306){\usebox{\plotpoint}}
\put(393,315){\usebox{\plotpoint}}
\put(410,327){\usebox{\plotpoint}}
\put(393,315){\usebox{\plotpoint}}
\put(410,179){\usebox{\plotpoint}}
\put(413,182){\usebox{\plotpoint}}
\put(417,185){\usebox{\plotpoint}}
\put(413,182){\usebox{\plotpoint}}
\put(417,185){\usebox{\plotpoint}}
\put(432,195){\usebox{\plotpoint}}
\put(448,200){\usebox{\plotpoint}}
\put(432,195){\usebox{\plotpoint}}
\put(410,327){\usebox{\plotpoint}}
\put(411,327){\usebox{\plotpoint}}
\put(412,328){\usebox{\plotpoint}}
\put(411,327){\usebox{\plotpoint}}
\put(412,328){\usebox{\plotpoint}}
\put(434,337){\usebox{\plotpoint}}
\put(448,347){\usebox{\plotpoint}}
\put(434,337){\usebox{\plotpoint}}
\put(448,200){\usebox{\plotpoint}}
\put(453,205){\usebox{\plotpoint}}
\put(460,209){\usebox{\plotpoint}}
\put(453,205){\usebox{\plotpoint}}
\put(460,209){\usebox{\plotpoint}}
\put(472,217){\usebox{\plotpoint}}
\put(486,222){\usebox{\plotpoint}}
\put(472,217){\usebox{\plotpoint}}
\put(448,347){\usebox{\plotpoint}}
\put(453,349){\usebox{\plotpoint}}
\put(455,352){\usebox{\plotpoint}}
\put(453,349){\usebox{\plotpoint}}
\put(455,352){\usebox{\plotpoint}}
\put(474,359){\usebox{\plotpoint}}
\put(486,368){\usebox{\plotpoint}}
\put(474,359){\usebox{\plotpoint}}
\put(486,222){\usebox{\plotpoint}}
\put(493,228){\usebox{\plotpoint}}
\put(501,232){\usebox{\plotpoint}}
\put(493,228){\usebox{\plotpoint}}
\put(501,232){\usebox{\plotpoint}}
\put(512,240){\usebox{\plotpoint}}
\put(524,244){\usebox{\plotpoint}}
\put(512,240){\usebox{\plotpoint}}
\put(486,368){\usebox{\plotpoint}}
\put(493,371){\usebox{\plotpoint}}
\put(497,376){\usebox{\plotpoint}}
\put(493,371){\usebox{\plotpoint}}
\put(497,376){\usebox{\plotpoint}}
\put(513,382){\usebox{\plotpoint}}
\put(524,390){\usebox{\plotpoint}}
\put(513,382){\usebox{\plotpoint}}
\put(524,244){\usebox{\plotpoint}}
\put(532,251){\usebox{\plotpoint}}
\put(542,256){\usebox{\plotpoint}}
\put(532,251){\usebox{\plotpoint}}
\put(542,256){\usebox{\plotpoint}}
\put(551,263){\usebox{\plotpoint}}
\put(562,267){\usebox{\plotpoint}}
\put(551,263){\usebox{\plotpoint}}
\put(524,390){\usebox{\plotpoint}}
\put(533,394){\usebox{\plotpoint}}
\put(538,400){\usebox{\plotpoint}}
\put(533,394){\usebox{\plotpoint}}
\put(538,400){\usebox{\plotpoint}}
\put(553,405){\usebox{\plotpoint}}
\put(562,412){\usebox{\plotpoint}}
\put(553,405){\usebox{\plotpoint}}
\put(562,267){\usebox{\plotpoint}}
\put(571,274){\usebox{\plotpoint}}
\put(582,280){\usebox{\plotpoint}}
\put(571,274){\usebox{\plotpoint}}
\put(582,280){\usebox{\plotpoint}}
\put(590,286){\usebox{\plotpoint}}
\put(600,289){\usebox{\plotpoint}}
\put(590,286){\usebox{\plotpoint}}
\put(562,412){\usebox{\plotpoint}}
\put(572,417){\usebox{\plotpoint}}
\put(579,423){\usebox{\plotpoint}}
\put(572,417){\usebox{\plotpoint}}
\put(579,423){\usebox{\plotpoint}}
\put(591,429){\usebox{\plotpoint}}
\put(600,435){\usebox{\plotpoint}}
\put(591,429){\usebox{\plotpoint}}
\put(600,289){\usebox{\plotpoint}}
\put(610,298){\usebox{\plotpoint}}
\put(622,304){\usebox{\plotpoint}}
\put(610,298){\usebox{\plotpoint}}
\put(622,304){\usebox{\plotpoint}}
\put(629,309){\usebox{\plotpoint}}
\put(638,312){\usebox{\plotpoint}}
\put(629,309){\usebox{\plotpoint}}
\put(600,435){\usebox{\plotpoint}}
\put(611,440){\usebox{\plotpoint}}
\put(619,447){\usebox{\plotpoint}}
\put(611,440){\usebox{\plotpoint}}
\put(619,447){\usebox{\plotpoint}}
\put(630,452){\usebox{\plotpoint}}
\put(638,458){\usebox{\plotpoint}}
\put(630,452){\usebox{\plotpoint}}
\put(638,312){\usebox{\plotpoint}}
\put(648,321){\usebox{\plotpoint}}
\put(662,328){\usebox{\plotpoint}}
\put(648,321){\usebox{\plotpoint}}
\put(662,328){\usebox{\plotpoint}}
\put(668,333){\usebox{\plotpoint}}
\put(676,336){\usebox{\plotpoint}}
\put(668,333){\usebox{\plotpoint}}
\put(638,458){\usebox{\plotpoint}}
\put(650,463){\usebox{\plotpoint}}
\put(658,471){\usebox{\plotpoint}}
\put(650,463){\usebox{\plotpoint}}
\put(658,471){\usebox{\plotpoint}}
\put(669,476){\usebox{\plotpoint}}
\put(676,481){\usebox{\plotpoint}}
\put(669,476){\usebox{\plotpoint}}
\put(676,336){\usebox{\plotpoint}}
\put(687,345){\usebox{\plotpoint}}
\put(701,352){\usebox{\plotpoint}}
\put(687,345){\usebox{\plotpoint}}
\put(701,352){\usebox{\plotpoint}}
\put(707,356){\usebox{\plotpoint}}
\put(714,359){\usebox{\plotpoint}}
\put(707,356){\usebox{\plotpoint}}
\put(676,481){\usebox{\plotpoint}}
\put(689,487){\usebox{\plotpoint}}
\put(697,495){\usebox{\plotpoint}}
\put(689,487){\usebox{\plotpoint}}
\put(697,495){\usebox{\plotpoint}}
\put(707,499){\usebox{\plotpoint}}
\put(714,504){\usebox{\plotpoint}}
\put(707,499){\usebox{\plotpoint}}
\put(714,359){\usebox{\plotpoint}}
\put(725,368){\usebox{\plotpoint}}
\put(740,376){\usebox{\plotpoint}}
\put(725,368){\usebox{\plotpoint}}
\put(740,376){\usebox{\plotpoint}}
\put(745,380){\usebox{\plotpoint}}
\put(752,382){\usebox{\plotpoint}}
\put(745,380){\usebox{\plotpoint}}
\put(714,504){\usebox{\plotpoint}}
\put(728,510){\usebox{\plotpoint}}
\put(736,519){\usebox{\plotpoint}}
\put(728,510){\usebox{\plotpoint}}
\put(736,519){\usebox{\plotpoint}}
\put(746,523){\usebox{\plotpoint}}
\put(752,528){\usebox{\plotpoint}}
\put(746,523){\usebox{\plotpoint}}
\put(752,382){\usebox{\plotpoint}}
\put(764,392){\usebox{\plotpoint}}
\put(779,400){\usebox{\plotpoint}}
\put(764,392){\usebox{\plotpoint}}
\put(779,400){\usebox{\plotpoint}}
\put(784,403){\usebox{\plotpoint}}
\put(790,405){\usebox{\plotpoint}}
\put(784,403){\usebox{\plotpoint}}
\put(752,528){\usebox{\plotpoint}}
\put(766,534){\usebox{\plotpoint}}
\put(775,543){\usebox{\plotpoint}}
\put(766,534){\usebox{\plotpoint}}
\put(775,543){\usebox{\plotpoint}}
\put(784,547){\usebox{\plotpoint}}
\put(790,551){\usebox{\plotpoint}}
\put(784,547){\usebox{\plotpoint}}
\put(790,405){\usebox{\plotpoint}}
\put(802,416){\usebox{\plotpoint}}
\put(818,423){\usebox{\plotpoint}}
\put(802,416){\usebox{\plotpoint}}
\put(818,423){\usebox{\plotpoint}}
\put(822,427){\usebox{\plotpoint}}
\put(828,429){\usebox{\plotpoint}}
\put(822,427){\usebox{\plotpoint}}
\put(790,551){\usebox{\plotpoint}}
\put(804,558){\usebox{\plotpoint}}
\put(814,567){\usebox{\plotpoint}}
\put(804,558){\usebox{\plotpoint}}
\put(814,567){\usebox{\plotpoint}}
\put(822,570){\usebox{\plotpoint}}
\put(828,575){\usebox{\plotpoint}}
\put(822,570){\usebox{\plotpoint}}
\put(828,429){\usebox{\plotpoint}}
\put(840,439){\usebox{\plotpoint}}
\put(857,447){\usebox{\plotpoint}}
\put(840,439){\usebox{\plotpoint}}
\put(857,447){\usebox{\plotpoint}}
\put(861,450){\usebox{\plotpoint}}
\put(866,452){\usebox{\plotpoint}}
\put(861,450){\usebox{\plotpoint}}
\put(828,575){\usebox{\plotpoint}}
\put(843,581){\usebox{\plotpoint}}
\put(853,591){\usebox{\plotpoint}}
\put(843,581){\usebox{\plotpoint}}
\put(853,591){\usebox{\plotpoint}}
\put(861,594){\usebox{\plotpoint}}
\put(866,598){\usebox{\plotpoint}}
\put(861,594){\usebox{\plotpoint}}
\put(866,452){\usebox{\plotpoint}}
\put(879,463){\usebox{\plotpoint}}
\put(896,471){\usebox{\plotpoint}}
\put(879,463){\usebox{\plotpoint}}
\put(896,471){\usebox{\plotpoint}}
\put(899,474){\usebox{\plotpoint}}
\put(904,476){\usebox{\plotpoint}}
\put(899,474){\usebox{\plotpoint}}
\put(866,598){\usebox{\plotpoint}}
\put(881,605){\usebox{\plotpoint}}
\put(891,614){\usebox{\plotpoint}}
\put(881,605){\usebox{\plotpoint}}
\put(891,614){\usebox{\plotpoint}}
\put(899,618){\usebox{\plotpoint}}
\put(904,622){\usebox{\plotpoint}}
\put(899,618){\usebox{\plotpoint}}
\put(904,476){\usebox{\plotpoint}}
\put(917,487){\usebox{\plotpoint}}
\put(934,495){\usebox{\plotpoint}}
\put(917,487){\usebox{\plotpoint}}
\put(934,495){\usebox{\plotpoint}}
\put(938,498){\usebox{\plotpoint}}
\put(942,499){\usebox{\plotpoint}}
\put(938,498){\usebox{\plotpoint}}
\put(904,622){\usebox{\plotpoint}}
\put(920,628){\usebox{\plotpoint}}
\put(930,638){\usebox{\plotpoint}}
\put(920,628){\usebox{\plotpoint}}
\put(930,638){\usebox{\plotpoint}}
\put(937,641){\usebox{\plotpoint}}
\put(942,645){\usebox{\plotpoint}}
\put(937,641){\usebox{\plotpoint}}
\put(942,499){\usebox{\plotpoint}}
\put(955,511){\usebox{\plotpoint}}
\put(973,519){\usebox{\plotpoint}}
\put(955,511){\usebox{\plotpoint}}
\put(973,519){\usebox{\plotpoint}}
\put(976,521){\usebox{\plotpoint}}
\put(980,523){\usebox{\plotpoint}}
\put(976,521){\usebox{\plotpoint}}
\put(942,645){\usebox{\plotpoint}}
\put(958,652){\usebox{\plotpoint}}
\put(969,662){\usebox{\plotpoint}}
\put(958,652){\usebox{\plotpoint}}
\put(969,662){\usebox{\plotpoint}}
\put(975,665){\usebox{\plotpoint}}
\put(980,669){\usebox{\plotpoint}}
\put(975,665){\usebox{\plotpoint}}
\put(980,523){\usebox{\plotpoint}}
\put(993,534){\usebox{\plotpoint}}
\put(1011,543){\usebox{\plotpoint}}
\put(993,534){\usebox{\plotpoint}}
\put(1011,543){\usebox{\plotpoint}}
\put(1014,545){\usebox{\plotpoint}}
\put(1018,546){\usebox{\plotpoint}}
\put(1014,545){\usebox{\plotpoint}}
\put(980,669){\usebox{\plotpoint}}
\put(996,676){\usebox{\plotpoint}}
\put(1007,686){\usebox{\plotpoint}}
\put(996,676){\usebox{\plotpoint}}
\put(1007,686){\usebox{\plotpoint}}
\put(1013,689){\usebox{\plotpoint}}
\put(1018,692){\usebox{\plotpoint}}
\put(1013,689){\usebox{\plotpoint}}
\put(1018,546){\usebox{\plotpoint}}
\put(1032,558){\usebox{\plotpoint}}
\put(1050,567){\usebox{\plotpoint}}
\put(1032,558){\usebox{\plotpoint}}
\put(1050,567){\usebox{\plotpoint}}
\put(1052,569){\usebox{\plotpoint}}
\put(1056,570){\usebox{\plotpoint}}
\put(1052,569){\usebox{\plotpoint}}
\put(1018,692){\usebox{\plotpoint}}
\put(1034,700){\usebox{\plotpoint}}
\put(1045,710){\usebox{\plotpoint}}
\put(1034,700){\usebox{\plotpoint}}
\put(1045,710){\usebox{\plotpoint}}
\put(1052,713){\usebox{\plotpoint}}
\put(1056,716){\usebox{\plotpoint}}
\put(1052,713){\usebox{\plotpoint}}
\put(1056,570){\usebox{\plotpoint}}
\put(1070,582){\usebox{\plotpoint}}
\put(1088,591){\usebox{\plotpoint}}
\put(1070,582){\usebox{\plotpoint}}
\put(1088,591){\usebox{\plotpoint}}
\put(1091,593){\usebox{\plotpoint}}
\put(1094,594){\usebox{\plotpoint}}
\put(1091,593){\usebox{\plotpoint}}
\put(1056,716){\usebox{\plotpoint}}
\put(1072,723){\usebox{\plotpoint}}
\put(1084,734){\usebox{\plotpoint}}
\put(1072,723){\usebox{\plotpoint}}
\put(1084,734){\usebox{\plotpoint}}
\put(1090,736){\usebox{\plotpoint}}
\put(1094,740){\usebox{\plotpoint}}
\put(1090,736){\usebox{\plotpoint}}
\put(1094,594){\usebox{\plotpoint}}
\put(1108,606){\usebox{\plotpoint}}
\put(1127,614){\usebox{\plotpoint}}
\put(1108,606){\usebox{\plotpoint}}
\put(1127,614){\usebox{\plotpoint}}
\put(1129,616){\usebox{\plotpoint}}
\put(1132,617){\usebox{\plotpoint}}
\put(1129,616){\usebox{\plotpoint}}
\put(1094,740){\usebox{\plotpoint}}
\put(1111,747){\usebox{\plotpoint}}
\put(1122,758){\usebox{\plotpoint}}
\put(1111,747){\usebox{\plotpoint}}
\put(1122,758){\usebox{\plotpoint}}
\put(1128,760){\usebox{\plotpoint}}
\put(1132,763){\usebox{\plotpoint}}
\put(1128,760){\usebox{\plotpoint}}
\put(1132,617){\usebox{\plotpoint}}
\put(1146,629){\usebox{\plotpoint}}
\put(1165,638){\usebox{\plotpoint}}
\put(1146,629){\usebox{\plotpoint}}
\put(1165,638){\usebox{\plotpoint}}
\put(1167,640){\usebox{\plotpoint}}
\put(1170,641){\usebox{\plotpoint}}
\put(1167,640){\usebox{\plotpoint}}
\put(1132,763){\usebox{\plotpoint}}
\put(1149,771){\usebox{\plotpoint}}
\put(1160,782){\usebox{\plotpoint}}
\put(1149,771){\usebox{\plotpoint}}
\put(1160,782){\usebox{\plotpoint}}
\put(1166,784){\usebox{\plotpoint}}
\put(1170,787){\usebox{\plotpoint}}
\put(1166,784){\usebox{\plotpoint}}
\put(1170,641){\usebox{\plotpoint}}
\put(1184,653){\usebox{\plotpoint}}
\put(1203,662){\usebox{\plotpoint}}
\put(1184,653){\usebox{\plotpoint}}
\put(1203,662){\usebox{\plotpoint}}
\put(1205,664){\usebox{\plotpoint}}
\put(1208,665){\usebox{\plotpoint}}
\put(1205,664){\usebox{\plotpoint}}
\put(1170,787){\usebox{\plotpoint}}
\put(1187,795){\usebox{\plotpoint}}
\put(1199,805){\usebox{\plotpoint}}
\put(1187,795){\usebox{\plotpoint}}
\put(1199,805){\usebox{\plotpoint}}
\put(1204,808){\usebox{\plotpoint}}
\put(1208,811){\usebox{\plotpoint}}
\put(1204,808){\usebox{\plotpoint}}
\put(1208,665){\usebox{\plotpoint}}
\put(1222,677){\usebox{\plotpoint}}
\put(1242,686){\usebox{\plotpoint}}
\put(1222,677){\usebox{\plotpoint}}
\put(1242,686){\usebox{\plotpoint}}
\put(1244,688){\usebox{\plotpoint}}
\put(1246,688){\usebox{\plotpoint}}
\put(1244,688){\usebox{\plotpoint}}
\put(1208,811){\usebox{\plotpoint}}
\put(1225,819){\usebox{\plotpoint}}
\put(1237,829){\usebox{\plotpoint}}
\put(1225,819){\usebox{\plotpoint}}
\put(1237,829){\usebox{\plotpoint}}
\put(1242,832){\usebox{\plotpoint}}
\put(1246,835){\usebox{\plotpoint}}
\put(1242,832){\usebox{\plotpoint}}
\put(1246,688){\usebox{\plotpoint}}
\put(1260,701){\usebox{\plotpoint}}
\put(1280,710){\usebox{\plotpoint}}
\put(1260,701){\usebox{\plotpoint}}
\put(1280,710){\usebox{\plotpoint}}
\put(1282,711){\usebox{\plotpoint}}
\put(1284,712){\usebox{\plotpoint}}
\put(1282,711){\usebox{\plotpoint}}
\put(1246,835){\usebox{\plotpoint}}
\put(1263,842){\usebox{\plotpoint}}
\put(1275,853){\usebox{\plotpoint}}
\put(1263,842){\usebox{\plotpoint}}
\put(1275,853){\usebox{\plotpoint}}
\put(1280,855){\usebox{\plotpoint}}
\put(1284,858){\usebox{\plotpoint}}
\put(1280,855){\usebox{\plotpoint}}
\put(1284,712){\usebox{\plotpoint}}
\put(1298,725){\usebox{\plotpoint}}
\put(1318,734){\usebox{\plotpoint}}
\put(1298,725){\usebox{\plotpoint}}
\put(1318,734){\usebox{\plotpoint}}
\put(1320,735){\usebox{\plotpoint}}
\put(1322,736){\usebox{\plotpoint}}
\put(1320,735){\usebox{\plotpoint}}
\put(1284,858){\usebox{\plotpoint}}
\put(1301,866){\usebox{\plotpoint}}
\put(1313,877){\usebox{\plotpoint}}
\multiput(1307.43,874.68)(-0.495,-0.489){14}{\rule{1.341pt}{0.118pt}}
\multiput(1310.22,874.92)(-9.217,-11.000){2}{\rule{0.670pt}{1.000pt}}
\put(1322,736){\usebox{\plotpoint}}
\put(1337,748){\usebox{\plotpoint}}
\put(1357,758){\usebox{\plotpoint}}
\put(1337,748){\usebox{\plotpoint}}
\put(1357,758){\usebox{\plotpoint}}
\put(1358,759){\usebox{\plotpoint}}
\put(1360,760){\usebox{\plotpoint}}
\put(1358,759){\usebox{\plotpoint}}
\put(1360,760){\usebox{\plotpoint}}
\put(1375,772){\usebox{\plotpoint}}
\put(1395,782){\usebox{\plotpoint}}
\put(1375,772){\usebox{\plotpoint}}
\put(1395,782){\usebox{\plotpoint}}
\put(1396,783){\usebox{\plotpoint}}
\put(1398,783){\usebox{\plotpoint}}
\put(1396,783){\usebox{\plotpoint}}
\put(1398,783){\usebox{\plotpoint}}
\put(1413,796){\usebox{\plotpoint}}
\put(1433,805){\usebox{\plotpoint}}
\put(1413,796){\usebox{\plotpoint}}
\put(1433,805){\usebox{\plotpoint}}
\put(1434,806){\usebox{\plotpoint}}
\put(1436,807){\usebox{\plotpoint}}
\put(1434,804.42){\rule{0.482pt}{1.000pt}}
\multiput(1435.00,804.92)(-1.000,-1.000){2}{\rule{0.241pt}{1.000pt}}
\end{picture}
\begin{center}
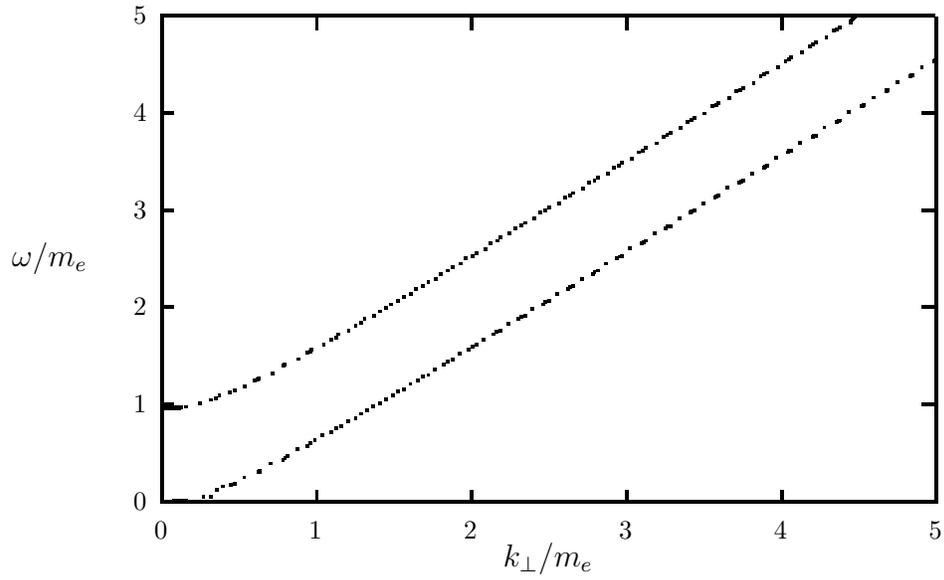
{\it Fig 2. Curve of propagation of neutrino normal to the magnetic field}
\end{center}
\end{document}